\documentclass[useAMS,usenatbib]{mn2e}

\usepackage{amssymb,amsmath}
\usepackage{longtable}
\usepackage{dcolumn}
\usepackage{graphicx}
\title[correlations between broad-line and jet emission variations for AGNs]
{Revisiting correlations between broad-line and jet emission
variations for AGNs: 3C 120 and 3C 273}

\author[H. T. Liu, J. M. Bai, H. C. Feng and S. K. Li]
{H. T. Liu$^{1,3}$\thanks{E-mail:
htliu@ynao.ac.cn}, J. M. Bai$^{1,3}$, H. C.
Feng$^{1,2,3}$ and S. K. Li$^{1,3}$ \\
$^{1}$Yunnan Observatories, Chinese Academy of
Sciences, Kunming, Yunnan 650011, China\\
$^{2}$University of Chinese Academy of Sciences,
Beijing 100049, China\\
$^{3}$Key Laboratory for the Structure and Evolution of Celestial
Objects, Chinese Academy of Sciences, Kunming, Yunnan 650011,
China}

\begin{document}

\date{Accepted . Received}


\maketitle

\label{firstpage}

\begin{abstract}
We restudy the issue of cross-correlations between broad-line and
jet emission variations, and aim to locate the position of radio
(and gamma-ray) emitting region in jet of active galactic nuclei
(AGNs). Considering the radial profiles of the radius and number
density of clouds in a spherical broad-line region (BLR), we
derive new formulae connecting jet emitting position
$R_{\rm{jet}}$ to time lag $\tau_{\rm{ob}}$ between broad-line and
jet emission variations, and BLR radius. Also, formulae are derived for a
disk-like BLR and a spherical shell BLR. The model-independent
FR/RSS method is used to estimate $\tau_{\rm{ob}}$. For 3C 120,
positive lags of about 0.3 yr are found between the 15 GHz
emission and the H$\beta$, H$\gamma$ and He II $\lambda 4686$
lines, including broad-line data in a newly published paper,
indicating the line variations lead the 15 GHz ones. Each of the
broad-line light curves corresponds to a radio outburst.
$R_{\rm{jet}}=$1.1--1.5 parsec (pc) are obtained for 3C 120. For
3C 273, a common feature of negative time lags is found in the
cross-correlation functions between light curves of radio emission
and the Balmer lines, and as well Ly$\alpha \/\ \lambda 1216$ and
C~{\sc iv} $\lambda 1549$ lines. $R_{\rm{jet}}=$ 1.0--2.6 pc are
obtained for 3C 273. The estimated $R_{\rm{jet}}$ are comparable
for 3C 120 and 3C 273, and the gamma-ray emitting positions will
be within $\sim$ 1--3 pc from the central engines. Comparisons
show that the cloud number density and radius radial distributions
and the BLR structures only have negligible effects on
$R_{\rm{jet}}$.
\end{abstract}

\begin{keywords}
galaxies: active -- galaxies: individual (3C 120, 3C 273) --
galaxies: jets -- quasars: emission lines -- radio continuum:
galaxies.
\end{keywords}

\section{INTRODUCTION}
According to the reverberation mapping model \citep[e.g.][]{BM82},
the broad emission line variations follow the ionizing continuum
variations through the photoionization process. The variability
correlations between the continuum and broad emission lines of
active galactic nuclei (AGNs) have been studied over the last
decades \citep[e.g.][]{Ka99,Ka00,Pe05}. A review about the
reverberation mapping researches is given in \citet[][and
references therein]{Ga09}. The jets can be ejected from inner
accretion disk in the vicinity of black hole
\citep[e.g.][]{Pe69,BZ77,BP82,Me01}. The disturbances in the
central engine are likely propagated outwards along the jets.
Observations show that dips in the X-ray emission, generated in
the central engine, are followed by ejections of bright
superluminal radio knots in the jets of AGNs and microquasars
\citep[e.g.][]{Ma02,Ar10}. The dips in the X-ray emission are well
correlated with the ejections of bright superluminal knots in the
radio jets of  3C 120 \citep{Ch09} and 3C 111 \citep{Ch11}. The
outbursts are physically linked to the ejections of superluminal
knots \citep[e.g.][]{Tu00}. These outbursts of broad-line and jet
emission should respond to the stronger disturbances in the
central engine. Cross-correlations between broad-line and jet
emission variations are expected, and a new method was proposed to
explain these correlations and constrain the positions of radio
and gamma-ray emitting regions \citep{Li11a}.

A ring broad-line region (BLR) is assumed to be perpendicular to
the jet axis (Paper I). This ring configuration of BLR is a toy
model relative to a disk-like BLR. Several groups find evidence
for disk-like BLRs \citep[e.g.][and references
therein]{Ko03,Nu13}. A necessary requirement of this disk-like BLR
is to have good variation features in the light curves. This is
especially the case for 3C120, as reported in \citet{Nu14}. The
sharp variation features are present in both the AGN continuum and
the BLR echo, and evidence for a nearly face-on disk-like BLR
geometry with an inclination of 10 degrees has been found for
3C120. The disk-like BLR is also used to explain the double-peaked
broad-lines in some AGNs, e.g. 3C 390.3 \citep{Zh13}. A spherical
BLR with some thickness is a widely used configuration for the
researches of the broad-line variability in AGNs
\citep[e.g.][]{Ka99}. The spherical BLR is also used to study
gamma-ray emission of AGNs \citep[e.g.][]{BLM09,LB06,Li08,Ta09}. A
spherical shell with a zero-thickness is used to study the origin
of gamma-ray emission of blazars \citep{Gh96}. The spherical BLR
consists of clouds with radial number density and radius pow-law
profiles of $n_{\rm{c}}(r)\propto r^{-p}$ and
$r_{\rm{c}}(r)\propto r^{q}$, respectively, where  $r$ is the
distance from the central engine to a cloud, and the pow-law
indexes $p$ and $q$ are positive \citep[e.g.][]{Ka99}. This
spherical BLR cloud model can fit variable broad emission lines in
AGNs. The BLR clouds may be bloated stars with extended envelopes,
and the emission-line intensities, profiles, and variability can
be fitted to the mean observed AGN spectrum under this model
\citep{AN94,AN97}. These two BLR models have a good agreement in
the trends of number density and density radial profiles of clouds
\citep[see][]{Ka99}.

3C 120 and 3C 111 are classified into the misaligned AGNs in the
third catalog of AGNs detected by \textit{Fermi}-LAT \citep{Ac15}.
The initial detection of 3C 120 with \textit{Fermi}-LAT was
reported by \citet{AAA10c}. Over the past 2 yr, \textit{Fermi}-LAT
sporadically detected 3C 120 with high significance in the MeV/GeV
band \citep{Ta15}. \citet{Ka11} argued that the gamma-ray emission
of broad-line radio galaxies detected by \textit{Fermi}-LAT are
most likely produced in the inner nucleus jets rather than large
scale jet structures. Broad-line blazar 3C 273 is brighter one of
\textit{Fermi}-LAT monitored sources\footnote
{http://fermi.gsfc.nasa.gov/ssc/}, and is included in the first
catalog of AGNs detected by \textit{Fermi}-LAT \citep{AAA10b}. The
gamma-ray flares detected with $\textit{Fermi}$-LAT for 3C 273
give a limit of gamma-ray emitting position smaller than 1.6
parsec (pc) from the central engine \citep{Ra13}. For 3C 120, it
was concluded that the gamma rays in the MeV/GeV band are more
favorably produced via the synchrotron self-Compton process,
rather than inverse Compton scattering of external photons coming
from BLR or dusty torus \citep{Ta15}. The conclusion is based on
their constraints on the relative positions of the gamma-ray and
radio emission regions. Therefore, the gamma-ray production
position is the key issue of how the gamma rays are produced.

The spherical BLR and the disk-like BLR are usually used to
produce the soft seed photons in the external Compton (EC) model
of gamma rays. The BLRs are important to gamma rays from blazars.
This importance arises from two factors. One is that the seed
photons from the BLR have significant influences on the EC
spectrum \citep[e.g.][]{Ta08,Le14b}. The other is photon-photon
absorption between the seed photons and the gamma-ray photons
\citep[see][]{Si94,Wa00,LB06,Li08,Si08,Ta08,BLM09,Ta09,Le14a}. The location
of gamma-ray emitting region relative to the BLR is the underlying
factor that controls how and how much the two factors influence
the gamma rays \citep{Li14}. For the disk-like BLR and the
spherical (shell) BLR with the same size, the relative positions
are different for the same gamma-ray emitting region in the jet.
In this paper, the spherical BLRs with different cloud
distributions will be focused on deriving new formulae to estimate
the radio emitting positions $R_{\rm{jet}}$ from the time lags
between variations of broad-lines and radio emission. As
comparison, we will estimate $R_{\rm{jet}}$ for the case of the
disk-like BLR, the spherical shell BLR and the ring BLR with a
zero-thickness. Once $R_{\rm{jet}}$ is known, the gamma-ray
emitting position $R_{\rm{\gamma}}$  could be constrained by
$R_{\rm{jet}}$ for AGNs.

The structure of this paper is as follows. Section 2 presents
method. Section 3 is for applications and contains three subsections:
subsection 3.1 presents analysis of time lag, subsection 3.2
application to 3C 120, and subsection 3.3 application to 3C 273.
section 4 is for variability amplitude.
Section 5 is for discussion and conclusions.

\section{METHOD}
In Paper I, the BLR is assumed to be a ring, and the plane of BLR
is assumed to be perpendicular to the jet axis. In this paper, we
assumed a spherical BLR, a BLR structure usually used. First, a
very thin spherical shell BLR is considered to get new equations.
Second, a spherical BLR is taken into account to get new formulae.
The geometrical structure is presented in Fig. 1 for the shell
BLR. First, the viewing angle to the jet axis is assumed to be
$\alpha=0$. As the disturbances from the central engine reach
point E (see Fig. 1a), where the jet emission are produced, i.e.
$R_{\rm{jet}}$=AE, the ionizing continuum photons travel from A to
C, and the line photons travel from C to D in time interval
$R_{\rm{jet}}/v_{\rm{d}}$. In the case, there is a zero-lag, and
we have $(R_{\rm{BLR}}+R_{\rm{jet}}-R_{\rm{BLR}}\cos
\theta)/c=R_{\rm{jet}}/v_{\rm{d}}$. Then we have for point C
\begin{equation}
R_{\rm{jet}}=\frac{R_{\rm{BLR}}(1-\cos
\theta)}{\frac{c}{v_{\rm{d}}}-1},
\end{equation}
where $\theta$ is the polar angle in spherical coordinates
$(r,\theta,\varphi)$, $R_{\rm{BLR}}$ is the BLR size, $v_{\rm{d}}$
is the travelling speed of disturbances down the jet and $c$ is
the speed of light.

If the disturbances reach point E and the line photons reach F,
the lines will lag the jet emission (see Fig. 1a). We have
$R_{\rm{BLR}}+R_{\rm{jet}}-R_{\rm{BLR}}\cos \theta -\tau c
=R_{\rm{jet}}/v_{\rm{d}}c$, and then
\begin{equation}
R_{\rm{jet}}=\frac{R_{\rm{BLR}}(1-\cos \theta)-\tau c
}{\frac{c}{v_{\rm{d}}}-1},
\end{equation}
where $\tau$ is the time lag of lines relative to the jet
emission. If the disturbances reach point E and the line photons
reach G, the lines will lead the jet emission (see Fig. 1a). We
have $R_{\rm{BLR}}+R_{\rm{jet}}-R_{\rm{BLR}}\cos \theta + \tau c
=R_{\rm{jet}}/v_{\rm{d}}c$, and then
\begin{equation}
R_{\rm{jet}}=\frac{R_{\rm{BLR}}(1-\cos \theta)+\tau c
}{\frac{c}{v_{\rm{d}}}-1},
\end{equation}
where $\tau$ is the time lag of the jet emission relative to the
lines.

Equations (1), (2), and (3) can be unified into
\begin{equation}
R_{\rm{jet}}=\frac{R_{\rm{BLR}}(1-\cos \theta)+\tau c
}{\frac{c}{v_{\rm{d}}}-1},
\end{equation}
where $\tau$ is zero, negative, and positive. As $\tau=0$,
equation (4) becomes equation (1). As $\tau<0$, equation (4)
becomes equation (2). As $\tau>0$, equation (4) becomes equation
(3). Equations (1)--(4) are obtained only for point C on spherical
shell of BLR (Fig. 1a). The observed line photons are from the
shell of BLR, and then the observed lag is an ensemble average
over all points of the shell. The surface element spanning from
$\theta$ to $\theta + d \theta$ and $\varphi$ to $\varphi +d
\varphi$ on a spherical surface at constant radius r is
$dS_{\rm{r}}=r^2\sin \theta d\theta d\varphi$. Thus the
differential solid angle is $d\Omega=dS_{\rm{r}}/r^2=\sin \theta
d\theta d\varphi$. Equation (4) can be integrated over every point
in the spherical surface of BLR, and then be averaged over $4\pi$.
Finally, we have
\begin{equation}
R_{\rm{jet}}=\frac{R_{\rm{BLR}}+\frac{\langle\tau_{\rm{ob}}
\rangle}{1+z} c }{\frac{c}{v_{\rm{d}}}-1},
\end{equation}
where $\langle\tau_{\rm{ob}} \rangle$ is the ensemble average of
$\tau_{\rm{ob}}$ over the spherical surface of BLR [and $\langle
1-\cos \theta \rangle
_{\rm{d\Omega}}=\int_{0}^{2\pi}\int_{0}^{\pi}(1-\cos \theta)\sin
\theta d\theta d\varphi /4\pi =1$], and $z$ is the redshift of
source. Equation (5) is same as equation (4) in Paper I.

Equations(5) is obtained for the viewing angle $\alpha=0$, which
is a special case. In general, $\alpha\neq 0$ (see Fig. 1b). As
the disturbances reach point B, where the jet emission are
generated, the ionizing continuum photons travel from point A to E
and the line photons travel from point E to G (assuming point D is
zero-lag point). Thus, we have
$R_{\rm{jet}}c/v_{\rm{d}}=\rm{AE+ED+DG}$, where
$\rm{AE}=R_{\rm{BLR}}$, $\rm{DG}=\tau c$, and
$\rm{ED}=\rm{FC}=\rm{AC-AF}=R_{\rm{jet}}\cos \alpha
-R_{\rm{BLR}}\cos \theta$ (see Fig. 1b). For point E in the
spherical surface of BLR, we have
$R_{\rm{jet}}c/v_{\rm{d}}=R_{\rm{jet}}\cos \alpha
-R_{\rm{BLR}}\cos \theta + R_{\rm{BLR}} + \tau c$, and then
\begin{equation}
R_{\rm{jet}}=\frac{R_{\rm{BLR}}(1-\cos \theta)+\tau
c}{\frac{c}{v_{\rm{d}}}-\cos \alpha},
\end{equation}
which becomes equation (4) as $\alpha =0$. Calculating the
ensemble average over $\theta$ and $\varphi$ in equation (6), we
have
\begin{equation}
R_{\rm{jet}}=\frac{R_{\rm{BLR}}+\frac{\langle\tau_{\rm{ob}}
\rangle}{1+z} c}{\frac{c}{v_{\rm{d}}}-\cos \alpha},
\end{equation}
which becomes equation (5) as $\alpha=0$, and is same as equation
(7) in Paper I. Here, $v_{\rm{d}}$ is equivalent to the bulk
velocity of jet $v_{\rm{j}}$, and $\langle \tau_{\rm{ob}}\rangle
\equiv \tau_{\rm{ob}}$ is the measured time lag of the jet
emission relative to the broad lines. From the velocity
$\beta=v_{\rm{j}}/c$ and the viewing angle $\alpha$, we have the
apparent speed $\beta_{\rm{a}}=\beta \rm{sin\alpha} /(1-\beta
\rm{cos\alpha})$, which gives $\beta =
\beta_{\rm{a}}/(\beta_{\rm{a}} \rm{cos\alpha}+\rm{sin\alpha})$.
Substituting the expression of $\beta$ for the velocity term in
equation (7), we have
\begin{equation}
R_{\rm{jet}}=\frac{\beta_{\rm{a}}}{\sin
\alpha}(R_{\rm{BLR}}+\frac{\langle\tau_{\rm{ob}} \rangle}{1+z} c),
\end{equation}
where $\beta_{\rm{a}}$, $\alpha$, and $R_{\rm{BLR}}$ are measured
from observations, and $\tau_{\rm{ob}}$ can be derived from the
cross-correlations between the broad-line and jet emission light
curves. On the other hand, if we have $R_{\rm{jet}}$, we can get
$R_{\rm{BLR}}$ from
\begin{equation}
R_{\rm{BLR}}=R_{\rm{jet}}\frac{\sin
\alpha}{\beta_{\rm{a}}}-\frac{\langle\tau_{\rm{ob}}
\rangle}{1+z}c.
\end{equation}

\begin{figure}
\begin{center}
\includegraphics[angle=-90,scale=0.3]{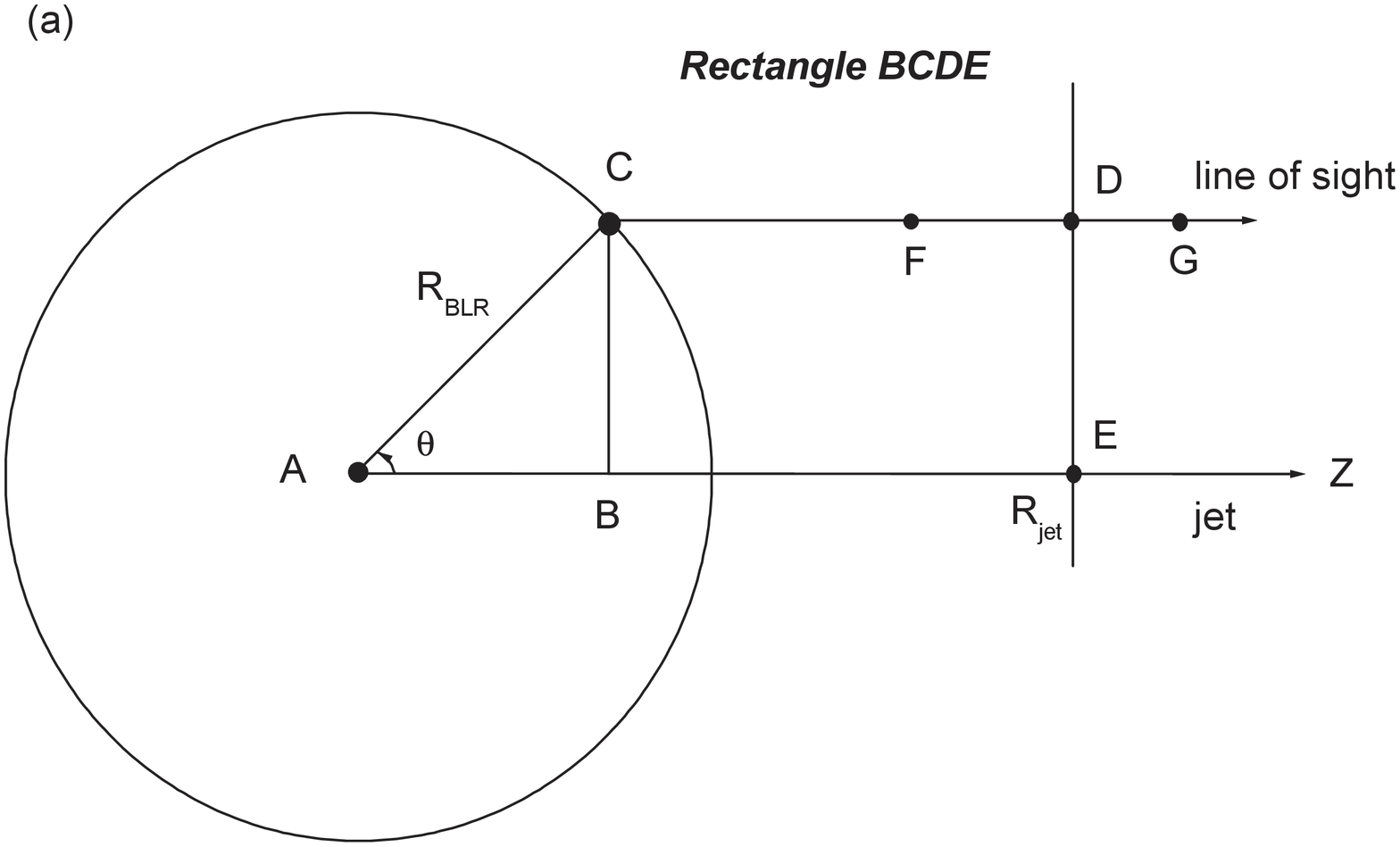}
\includegraphics[angle=-90,scale=0.3]{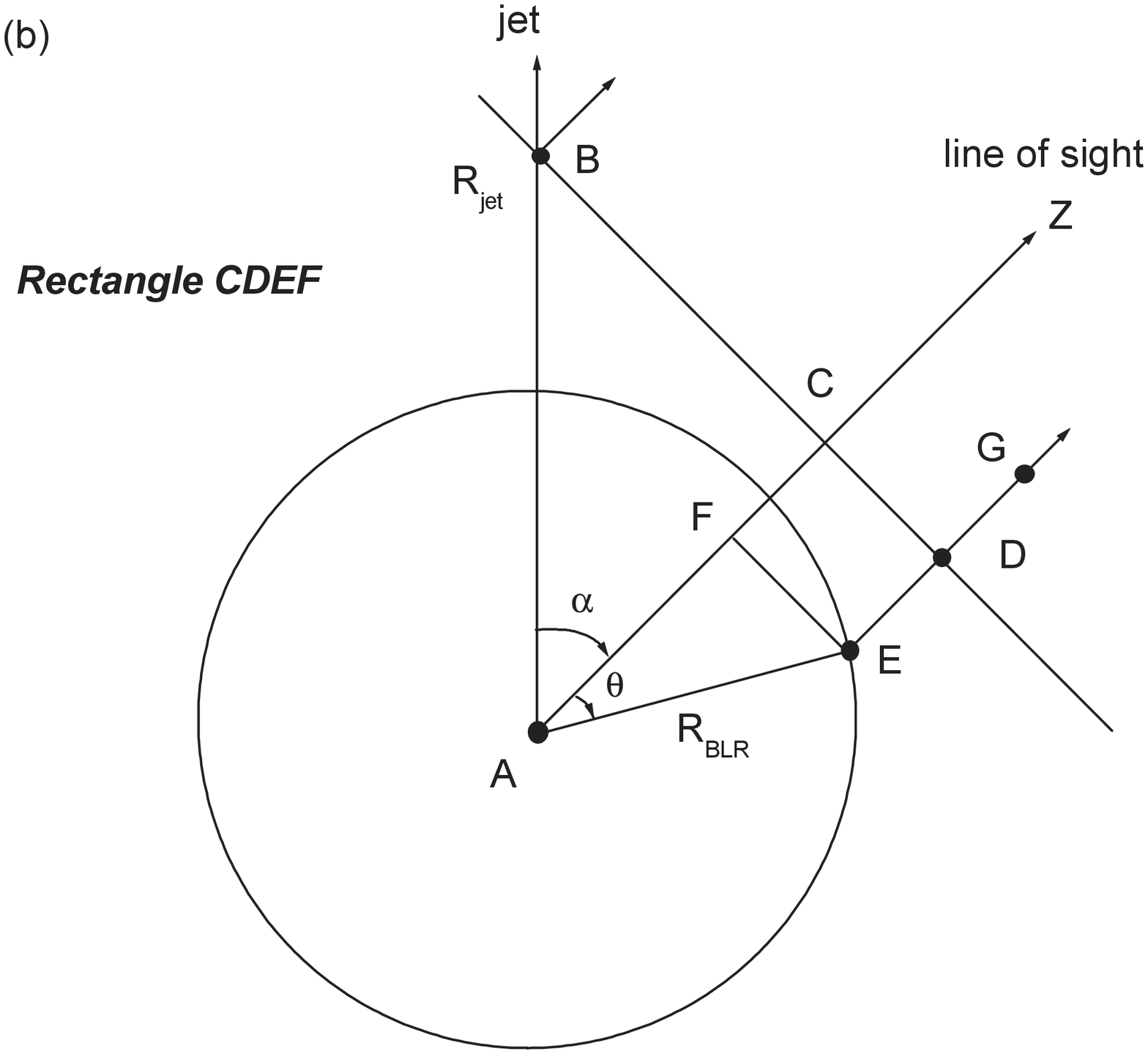}

\end{center}
 \caption{Sketch of axial cross section of the spherical geometry assumed, and it is similar to that used in the reverberation mapping
 method of broad emission lines. $R_{\rm{BLR}}$ is the size of BLR. ($a$) the angle between the line of
 sight and the jet axis $\alpha=0$. ($b$) $\alpha \neq 0$. }
  \label{fig1}
\end{figure}

These deduced formulae are based on the simplification of a
spherical shell BLR of zero thickness. In fact, the real BLRs have
some thickness that must be considered. The real BLRs consist of
many clouds (see Fig. 2). These clouds have the number density
$n_{\rm{c}}(r)$ and the cross section $\sigma_{\rm{c}}(r)$ at
radius $r$, respectively. For a thin spherical shell in the range
of $r \rightarrow r+dr$ (see Fig. 2), it has a covering factor of
$df_{\rm{cov}}(r)=n_{\rm{c}}(r)\sigma_{\rm{c}}(r) dr$ and a volume
$4\pi r^2 dr$. The emissivity (in $\rm{erg \/\ s^{-1} \/\ cm^{-3}
\/\ sr^{-1}}$), reprocessed due to ultraviolet (UV) radiation
luminosity $L_{\rm{UV}}$, inside the thin spherical shell of BLR
at the radius $r$ is
\begin{equation}
j_{\rm{BLR}}=\frac{L_{\rm{UV}}df_{\rm{cov}}(r)}{16\pi^2 r^2 dr}
=\frac{L_{\rm{UV}} n_{\rm{c}}(r) \sigma_{\rm{c}}(r)} {16\pi^2
r^2},
\end{equation}
where $\sigma_{\rm{c}}(r)=\pi r^2_{\rm{c}}$ and $r_{\rm{c}}$ is
the radius of clouds at the radius $r$. \citet{Ka99} showed the
power-law profiles of $r_{\rm{c}}(r)$ and $n_{\rm{c}}(r)$ as
$r_{\rm{c}}(r)=r_{\rm{c0}}(r/R_{\rm{BLR,in}})^{q}$ and
$n_{\rm{c}}(r)=n_{\rm{c0}}(r/R_{\rm{BLR,in}})^{-p}$ with
$n_{\rm{c0}}$ and $r_{\rm{c0}}$ to be the number density and the
radius of clouds at $R_{\rm{BLR,in}}$, respectively. Then we have
\begin{equation}
j_{\rm{BLR}} =\frac{L_{\rm{UV}} n_{\rm{c0}} r^2_{\rm{c0}}} {16\pi
R_{\rm{BLR,in}}^{2q-p} } r^{2q-p-2},
\end{equation}

Since the observed fluxes of broad emission lines are produced by
the entire BLR, the time lag $\langle\tau_{\rm{ob}} \rangle$ in
equation (8) will be a flux-weighted average value for all the
spherical shells in Fig. 2. Thus $R_{\rm{BLR}}$ in the right hand
of equation (8) will be replaced with a flux-weighted average
\begin{equation}
\begin{split}
\langle R_{\rm{BLR}} \rangle &=\frac{\int
^{R_{\rm{BLR,out}}}_{R_{\rm{BLR,in}}}r dF_{\rm{BLR}}(r)}{\int
^{R_{\rm{BLR,out}}}_{R_{\rm{BLR,in}}}dF_{\rm{BLR}}(r)} \\
&=\frac{\int^{R_{\rm{BLR,out}}}_{R_{\rm{BLR,in}}}j_{\rm{BLR}}(r)rdr}{\int
^{R_{\rm{BLR,out}}}_{R_{\rm{BLR,in}}}j_{\rm{BLR}}(r)dr}  \\
&=\frac{\int^{R_{\rm{BLR,out}}}_{R_{\rm{BLR,in}}}r^{2q-p-1}dr}{\int
^{R_{\rm{BLR,out}}}_{R_{\rm{BLR,in}}}r^{2q-p-2}dr},
\end{split}
\end{equation}
where $d F_{\rm{BLR}}(r)$ is the differential broad-line flux of
the thin spherical shell in the range of $r\rightarrow r+dr$, and
$d F_{\rm{BLR}}(r)=4\pi j_{\rm{BLR}}(r)dr$. Finally, we have
\begin{equation}
\begin{split}
R_{\rm{jet}}&=\frac{\beta_{\rm{a}}}{\sin \alpha} \left( \frac{\int
^{R_{\rm{BLR,out}}}_{R_{\rm{BLR,in}}}r^{2q-p-1}dr}{\int
^{R_{\rm{BLR,out}}}_{R_{\rm{BLR,in}}}r^{2q-p-2}dr}+\frac{\langle\tau_{\rm{ob}}
\rangle}{1+z} c \right) \\
&=\frac{\beta_{\rm{a}}}{\sin \alpha}
\left(\frac{2q-p-1}{2q-p}
\frac{R_{\rm{BLR,out}}^{2q-p}-{R_{\rm{BLR,in}}^{2q-p}}}{R_{\rm{BLR,out}}^{2q-p-1}-{R_{\rm{BLR,in}}^{2q-p-1}}}+\frac{\langle\tau_{\rm{ob}}
\rangle}{1+z} c \right),
\end{split}
\end{equation}
if $2q-p-1\neq -1$ and  $2q-p-2\neq -1$. \citet{Ka99} got $q=1/3$
and $p=3/2$ for the spherical BLR. Then equation (13) becomes
\begin{equation}
R_{\rm{jet}}=\frac{\beta_{\rm{a}}}{\sin \alpha} \left(\frac{11}{5}
\frac{R_{\rm{BLR,out}}^{-\frac{5}{6}}-{R_{\rm{BLR,in}}^{-\frac{5}{6}}}}{R_{\rm{BLR,out}}^{-\frac{11}{6}}-{R_{\rm{BLR,in}}^{-\frac{11}{6}}}}+\frac{\langle\tau_{\rm{ob}}
\rangle}{1+z} c \right).
\end{equation}

\begin{figure}
\begin{center}
\includegraphics[angle=-90,scale=0.3]{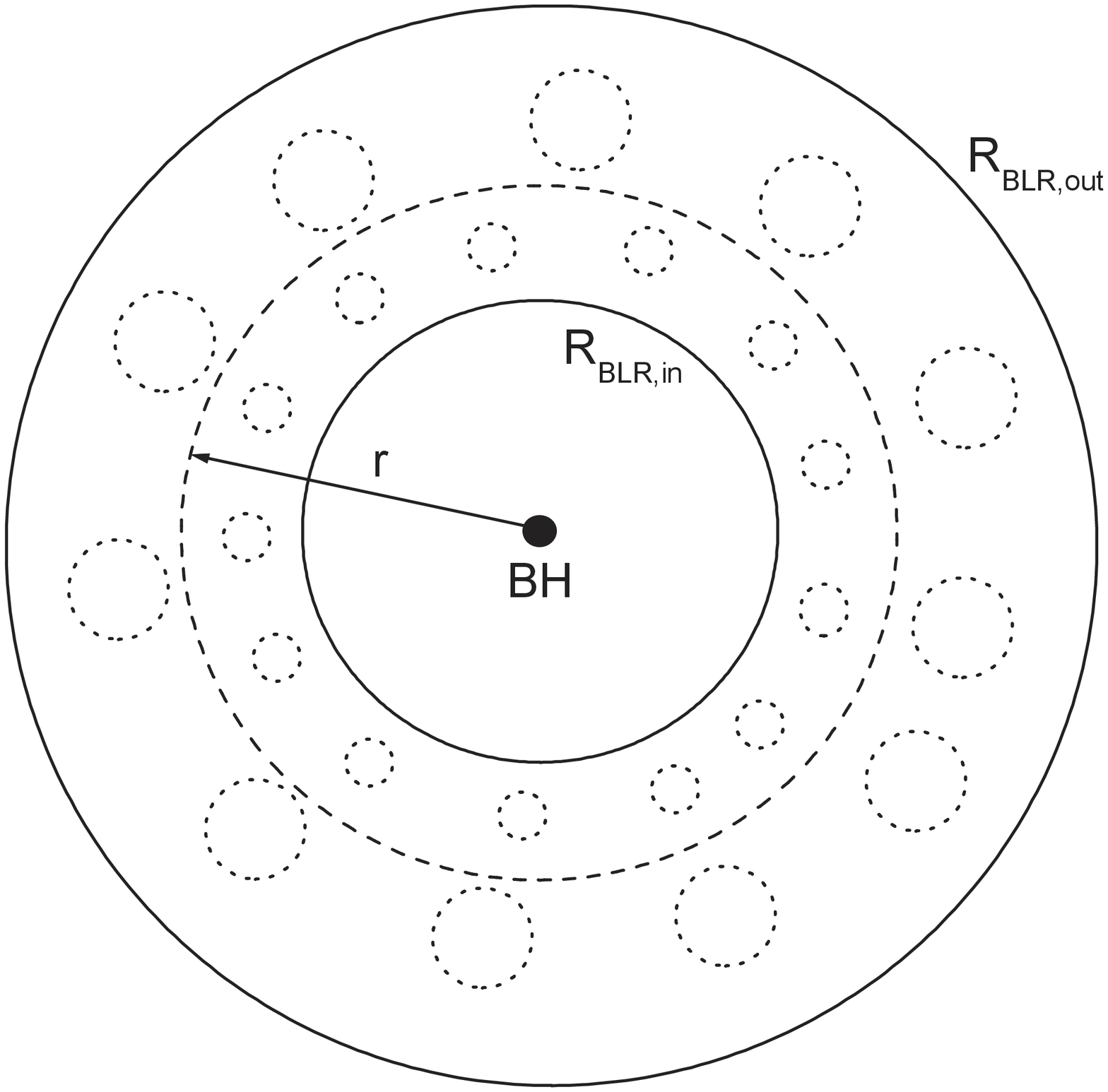}
\end{center}
 \caption{Sketch of axial cross section of the spherical geometry with some thickness and similar to Fig. 1 of Liu \& Bai (2006). }
  \label{fig2}
\end{figure}

\section{APPLICATIONS}
The new formulae are applied to broad-line radio galaxy 3C 120 at
$z=0.033$ and blazar 3C 273 at redshift $z=0.158$, which were
detected with \textit{Fermi}-LAT.

\subsection{Analysis of Time Lag}
The z-transformed discrete correlation function
\citep[ZDCF;][]{Al97} is used to analyze cross-correlation. The
centroid lag in cross-correlation function (CCF) is taken to
characterize the time lag between broad-line and jet emission
variations. The centroid time lag $\tau_{\rm{cent}}$ is computed
by all the points with correlation coefficients not less than 0.8
times the maximum of correlation coefficients in the CCF bumps
closer to the zero-lag \citep[see][]{Li11b}. The model-independent
flux randomization/random subset selection (FR/RSS) Monte Carlo
method \citep{Pe98b} is used to get the cross-correlation centroid
distributions (CCCDs). The averages of CCCDs are taken as the time
lags between the broad-line and jet emission variations, and the
standard deviations of the same CCCDs are adopted as our formal
$1\sigma$ uncertainties of these time lags. This treatment is same
as in \citet{Gr12} for the CCF analysis. Hereafter,
$\tau_{\rm{cent}}$ equivalent to $\tau_{\rm{ob}}$.

\subsection{3C 120}
The 15 GHz light curve is published in \citet{Ri11}, and is
obtained with a higher sampling of $\sim$ 60 times $\rm{yr^{-1}}$
by the OVRO 40 m blazar monitoring program\footnote
{http://www.astro.caltech.edu/ovroblazars}. The H$\beta$ line
light curves are from three different reverberation mapping
monitoring works \citep{Gr12,Nu12,Ko14}. The data of \citet{Gr12}
and \citet{Nu12} have a dense sampling of 20 times
$\rm{month^{-1}}$. The data of \citet{Ko14} published currently
have a rare sampling of 5 times $\rm{month^{-1}}$ and larger flux
errors compared to the other two works. The data numbers and
durations of light curves are presented in Table 1. These light
curves are presented in Fig. 3. The combined H$\beta$ light curves
of the three works are used to cross-correlate with the 15 GHz
light curve. There is a positive time lag (see Figs. 4a and 4b and
Table 2), which means the H$\beta$ line variations leading the 15
GHz variations. This new time lag is consistent with the previous
result of $\tau_{\rm{cent}}=0.34\pm 0.01$ yr in \citet{Li14}. For
the H$\beta$ line, there are two outbursts, the outburst in
\citet{Ko14} and the one in \citet{Gr12}, likely corresponding to
the radio Outbursts I and II, respectively (see Fig. 3). The
H$\gamma$ and He II $\lambda 4686$ outbursts in \citet{Ko14}
likely correspond to Outburst I.
\begin{figure}
 \begin{center}
 \includegraphics[angle=-90,scale=0.25]{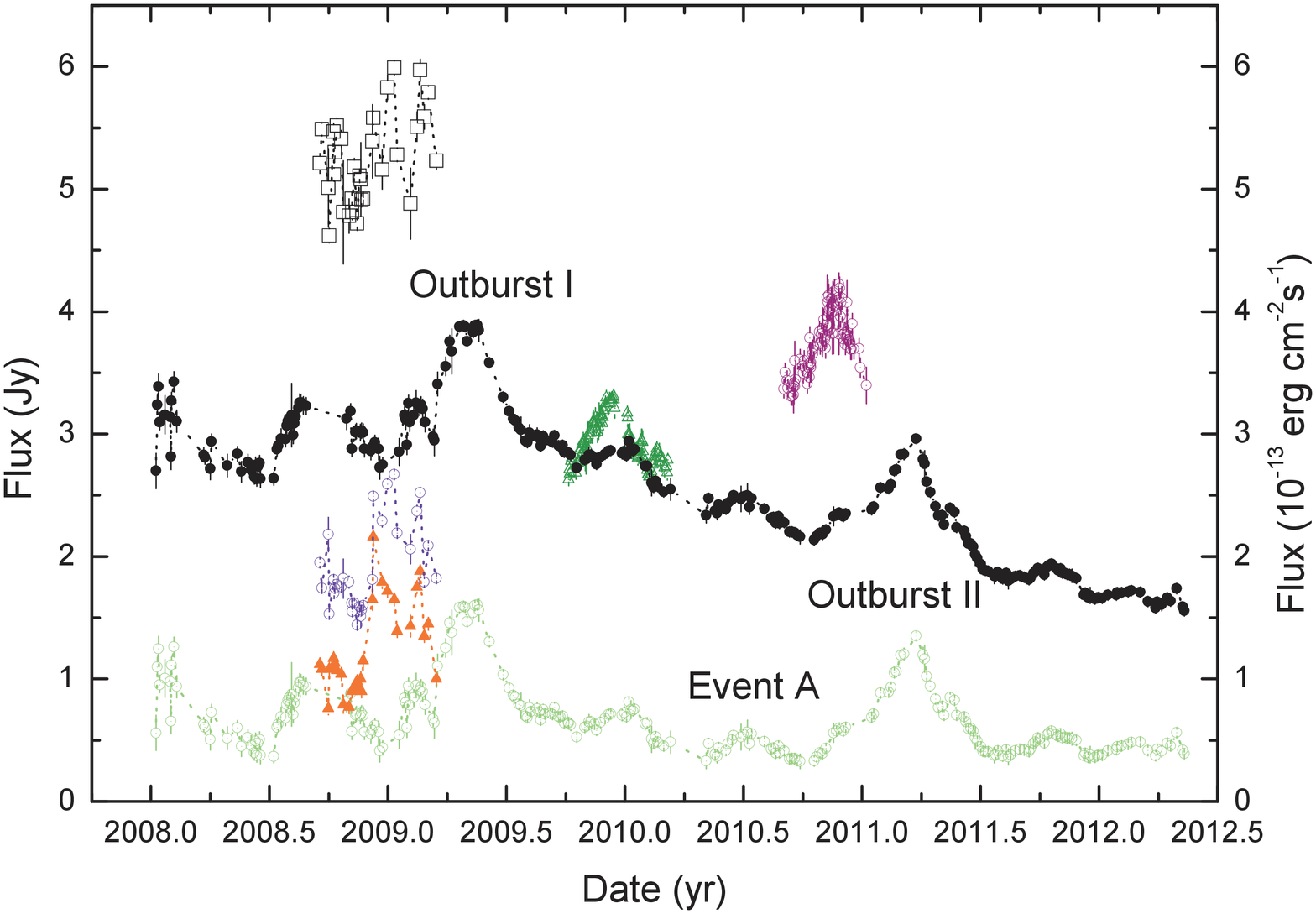}
 \end{center}
 \caption{Light curves of broad-lines and 15 GHz emission. Olive triangles denote the H$\beta$ light curve of \citet{Nu12}.
   Purple circles denote the H$\beta$ light curve of \citet{Gr12}. Squares denote the H$\beta$ light curve of \citet{Ko14}.
   Blue circles are the H$\gamma$ light curve of \citet{Ko14}.
   Red triangles denote the He II $\lambda 4686$ light curve of \citet{Ko14}. Black circles denote the 15 GHz light curve in units of Jy.
  The lines are in units of $10^{-13}\rm{\/\ erg \/\ cm^{-2}\/\ s^{-1}}$.
  Green circles are the 15 GHz light curve subtracted by an assumed simple baseline \citep{Li14}.}
  \label{fig3}
\end{figure}
\begin{figure}
 \begin{center}
 \includegraphics[angle=-90,scale=0.30]{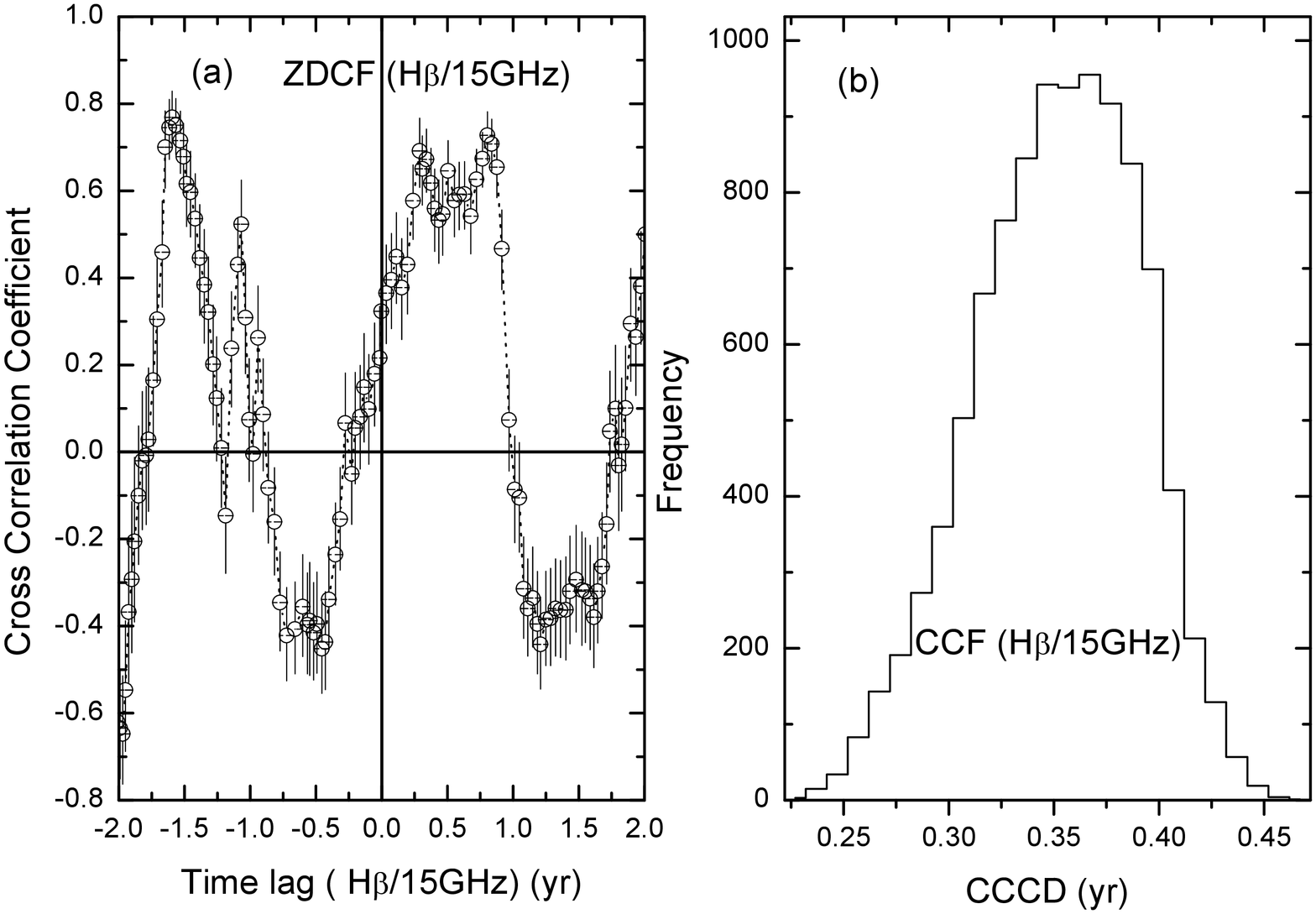}
 \includegraphics[angle=-90,scale=0.30]{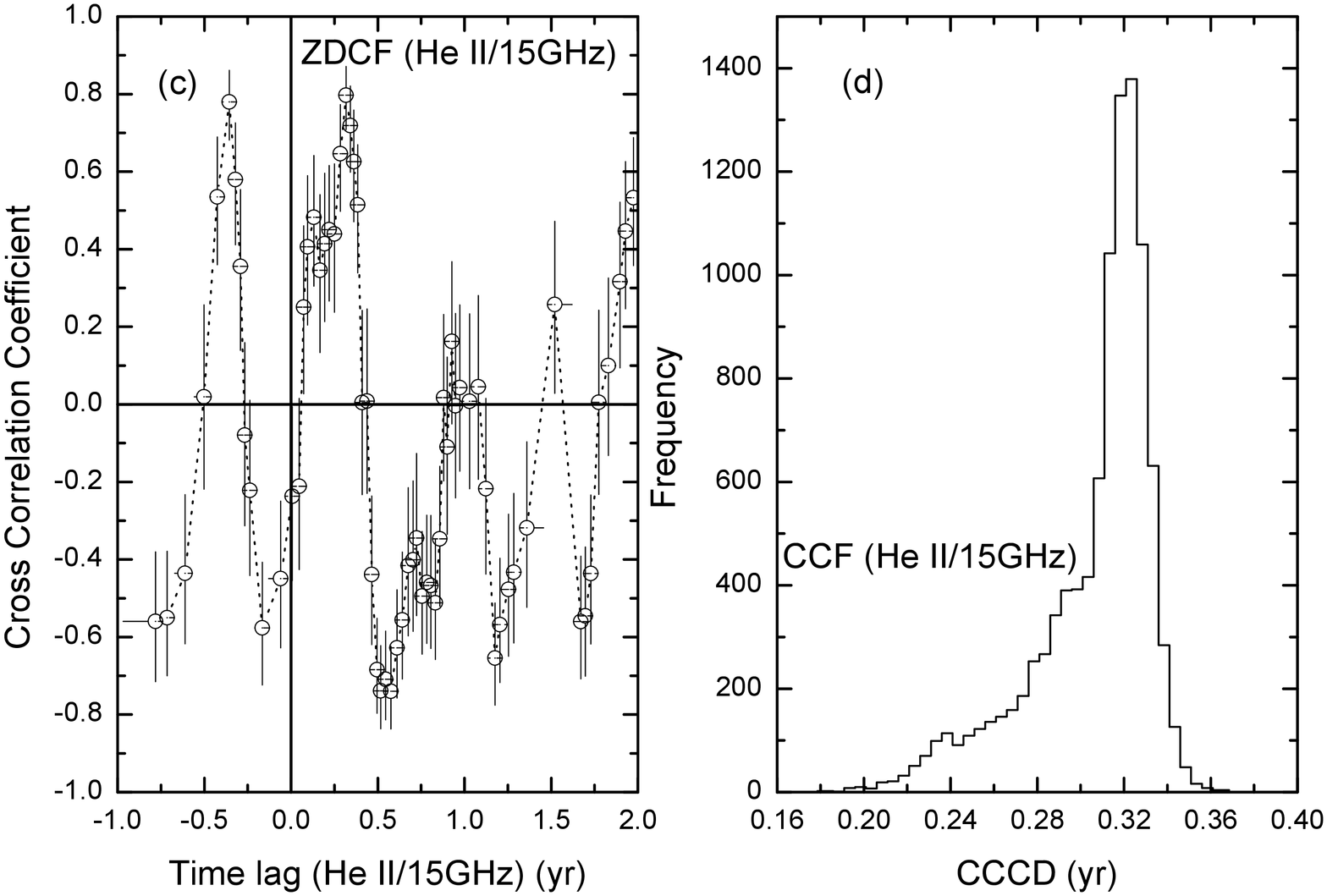}
  \includegraphics[angle=-90,scale=0.30]{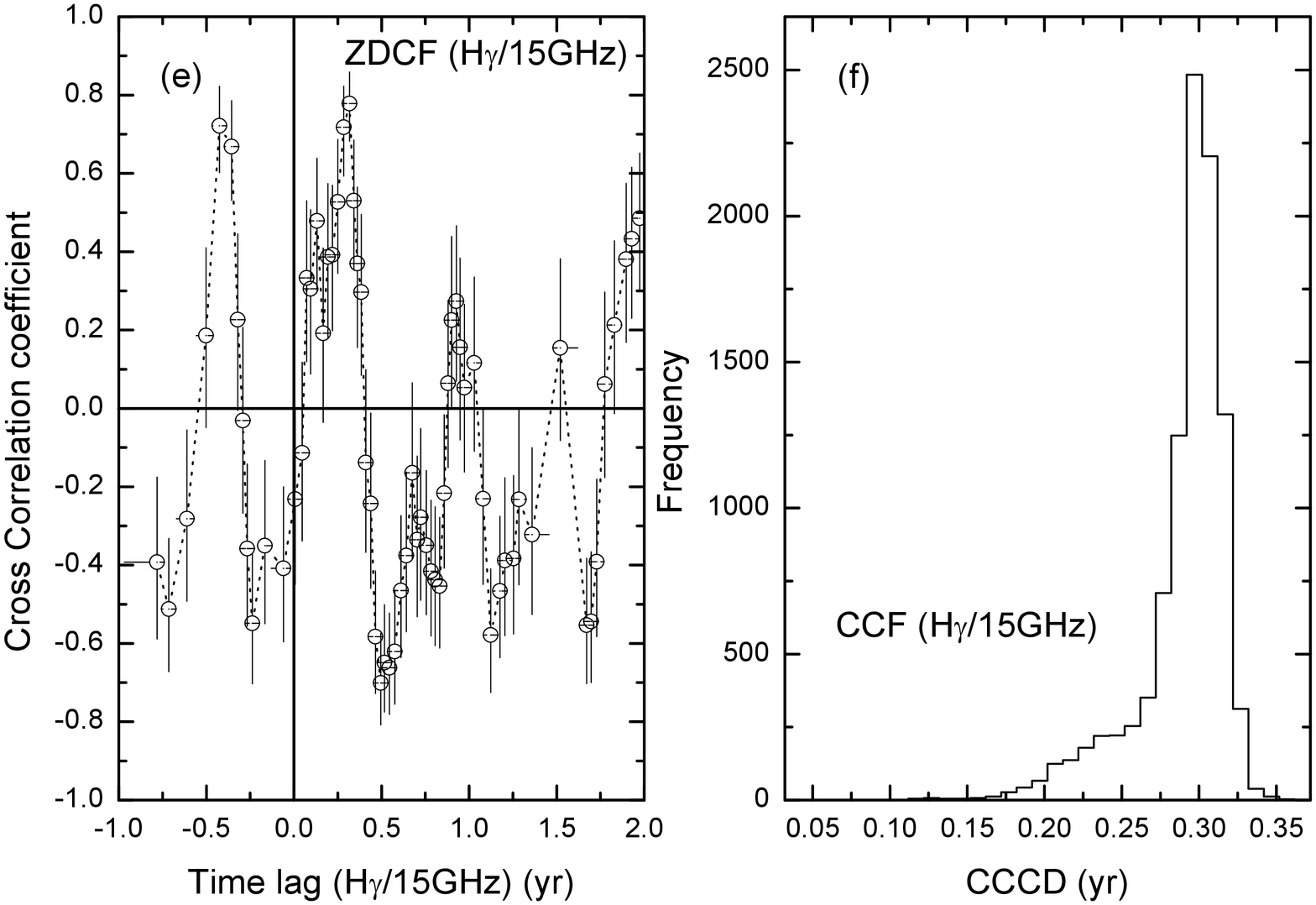}
 \end{center}
\caption{ZDCFs between the modified 15 GHz light curve and the
broad-line light curves, and CCCDs obtained with the FR/RSS method
in Monte Carlo simulations of 10,000 runs. (a) and (b) for the
total H$\beta$ light curve. (c) and (d) for the He II $\lambda
4686$ light curve. (e) and (f) for the H$\gamma$ light curve.}
\label{fig4}
\end{figure}
\begin{table}
\centering
\begin{minipage}{80mm}

\caption{Fractional variability of 3C 120 \label{tbl-1}}

\begin{tabular}{ccccc}

\hline\hline

Component & $F_{\rm{var}}$ & $\sigma_{\rm{F_{\rm{var}}}}$ & $N$ & $D$ (yr)\\
(1)&(2)&(3)&(4)&(5)\\

\hline

H$\beta$ $^{a}$ &0.06&0.01 & 31 & 0.49 \\
H$\beta$ $^{b}$ &0.057&0.005&102 & 0.42 \\
H$\beta$ $^{c}$ &0.058&0.005&85 & 0.35 \\
H$\beta$ $^{d}$ &0.23&0.01 &218 & 1.26 \\
H$\gamma$     &0.18 &0.02 &31 & 0.49 \\
He II $\lambda 4686$&0.30&0.04&31 & 0.49 \\
15 GHz        &0.22 &0.01 &257 & 4.34 \\
\hline

Outburst I       &0.34 &0.04 &45 & 0.66 $^\dag$\\
Outburst II       &0.27 &0.04 & 31 & 0.59 $^\dag$\\
Event A       &0.15 &0.03 &27 & 0.46 $^\dag$\\
\hline
H$\beta$ $^{b}$ &0.26&0.02&102 & 0.42 $^\dag$\\
H$\beta$ $^{c}$ &0.21&0.02&85 & 0.35 $^\dag$\\
H$\gamma$     &0.31 &0.04 &31 & 0.49 $^\dag$\\
He II $\lambda 4686$&0.30&0.04&31 & 0.49 $^\dag$\\

\hline
\end{tabular}
\\Notes: Component: different components; $F_{\rm{var}}$: Fractional variability of light curves;
$\sigma_{\rm{F_{\rm{var}}}}$: the error of $F_{\rm{var}}$; $N$:
the observational data numbers in the light curves; $D$: the
durations of light curves.
\\$^{a}$ the H$\beta$ light curve in \citet{Ko14}.
\\$^{b}$ the H$\beta$ light curve in \citet{Nu12}.
\\$^{c}$ the H$\beta$ light curve in \citet{Gr12}.
\\$^{d}$ the total H$\beta$ light curve of three periods in notes a, b, and c.
\\$^\dag$ the light curves denoted by the black solid and blue circles in Fig. 8.

\end{minipage}
\end{table}

The apparent speeds of the moving components with well-determined
motions are all within a range of $\beta_{\rm{a}}= 4.0 \pm 0.2$
for 3C 120 \citep[see][]{Ch09}. \citet{Gr12} obtained a new size
of $R_{\rm{BLR}}=27.2\pm 1.1$ light-days in their dense mapping
observations for the H$\beta$ line. \citet{Nu12} also obtained a
new BLR size with small errors, which is consistent with that in
\citet{Gr12}. This value of $R_{\rm{BLR}}=0.075\pm 0.003$ lt-yr is
adopted here. The $1\sigma$ upper and lower limits of
$R_{\rm{BLR}}$ are taken as $R_{\rm{BLR,out}}=0.078$ lt-yr and
$R_{\rm{BLR,in}}=0.072$ lt-yr, respectively. The 43 GHz VLBA
observations give the global parameters of the jet with a viewing
angle $\alpha = 20.5\pm 1.8^{\circ}$ for 3C 120 \citep{Jo05}. For
$\beta_{\rm{a}}=4.0\pm 0.2$, $\alpha = 20.5\pm 1.8^{\circ}$,
$\tau_{\rm{cent}}=0.35 \pm 0.04$ yr, $R_{\rm{BLR,out}}=0.078$ and
$R_{\rm{BLR,in}}=0.072$ lt-yr, we have $R_{\rm{jet}}=1.45 \pm
0.02$ pc from Monte Carlo simulations based on equation (14) (see
Table 2). This radio emitting region is at the pc-scale distance
from the central engine.
\begin{table*}
\centering
\begin{minipage}{130mm}
\caption{Estimated values of time lags $\tau_{\rm{cent}}$ and
$R_{\rm{jet}}$ for 3C 120 \label{tbl-2}}

\begin{tabular}{ccccccc}

\hline\hline

Lines & $R_{\rm{BLR}}$ (lt-yr)& Ref. & $\tau_{\rm{cent}}$ (yr) &
$R_{\rm{jet}}^{\rm{spher}}$ (pc) & $R_{\rm{jet}}^{\rm{disk}}$ (pc)
& $R_{\rm{jet}}^{\rm{shell}}$ (pc)  \\
(1)&(2)&(3)&(4)&(5)&(6)&(7)\\

\hline

H$\beta$&$0.075^{+0.003}_{-0.003}$&1&$0.35\pm 0.04$&$1.45\pm 0.02$&$1.45\pm 0.02$&$1.45\pm 0.02$\\
H$\gamma$&$0.065^{+0.013}_{-0.011}$&2&$0.29\pm 0.03$&$1.21\pm 0.01$&$1.21\pm 0.01$ &$1.21\pm 0.02$\\

He II $\lambda 4686$&$0.033^{+0.021}_{-0.019}$&2&$0.31\pm 0.03$&$1.13\pm 0.01$&$1.14\pm 0.01$&$1.17\pm 0.03$\\

\hline
\end{tabular}
\\Notes: Lines: line names; $R_{\rm{BLR}}$: BLR sizes,
$R_{\rm{BLR,in}}=R_{\rm{BLR}}-\sigma_{\rm{R_{\rm{BLR}}}}$ and
$R_{\rm{BLR,out}}=R_{\rm{BLR}}+\sigma_{\rm{R_{\rm{BLR}}}}$ are
taken to estimate $R_{\rm{jet}}$; Ref.: the references for column
2; $\tau_{\rm{cent}}$: Time lags, defined as $\tau_{\rm{cent}}=
t_{\rm{radio}}-t_{\rm{line}}$, between broad-lines and radio
emission; $R_{\rm{jet}}^{\rm{spher}}$: $R_{\rm{jet}}$ estimated
with equation (14) for the spherical BLR;
$R_{\rm{jet}}^{\rm{disk}}$: $R_{\rm{jet}}$ estimated with equation
(19) for the disk-like BLR; $R_{\rm{jet}}^{\rm{shell}}$:
$R_{\rm{jet}}$ estimated with equation (8) for the spherical shell
or ring BLR.\\
\textbf{References}: (1) \citealt{Gr12}; (2) \citealt{Ko14}.
\end{minipage}
\end{table*}

In \citet{Ko14}, the He II $\lambda 4686$ line was also monitored
in the reverberation mapping observations, and this line light
curve has a smaller flux errors than those in the H$\beta$ line
light curve (see Fig. 3). The He II $\lambda 4686$ line light
curve likely correspond to Outburst I in the 15 GHz light curve.
The ZDCF method and the FR/RSS method show a positive time lag of
the He II $\lambda 4686$ line relative to the 15 GHz variations
(see Figs. 4c and 4d and Table 2). The H$\gamma$ line variations
also seem to lead the 15 GHz variations by about 0.3 yr (see Fig.
3). Figs. 4e and 4f present the results from the ZDCF method and
the FR/RSS method. There is a positive time lag around 0.3 yr (see
Table 2 and Figs. 4e and 4f). These results indicate that the line
variations lead the 15 GHz variations by about 0.3 yr, and confirm
the positive lag from the H$\beta$ line. \citet{Ko14} estimated
the sizes of the He II $\lambda 4686$ and H$\gamma$ lines,
$R_{\rm{BLR}}=12.0^{+7.5}_{-7.0}$ and
$R_{\rm{BLR}}=23.9^{+4.6}_{-3.9}$ light-days, respectively. Thus
we have $R_{\rm{jet}}=1.13 \pm 0.01$ and $1.21 \pm 0.01$ pc from
the He II $\lambda 4686$ and H$\gamma$ lines, respectively (see
Table 2). The positions around one pc from the central engine are
obtained from the time lags between these variations of the 15 GHz
emission and the H$\beta$, H$\gamma$ and He II $\lambda 4686$
lines. These pc-scale distances from the central engine are larger
than the BLR sizes by about two orders of magnitude. This
indicates for 3C 120 that the gamma rays detected with
\textit{Fermi}-LAT are likely from the synchrotron-self Compton
(SSC) processes in the jet if the gamma-ray emitting regions are
around the radio emitting regions.

\subsection{3C 273}
From the 3C 273 database\footnote {http://isdc.unige.ch/3c273/}
hosted by the ISDC \citep{Tu99} and updated by \citet{So08}
(references therein), we take the 5, 8, 15, 22, and 37 GHz radio
light curves. The sampling rates of 5, 8, 15, 22 and 37 GHz are
29, 40, 40, 44 and 46 times per year for the data considered,
respectively. Only good data (Flag $>=$0) are used in the light
curves considered here. Light curves of broad lines H$\alpha$,
H$\beta$, and H$\gamma$ are from spectrophotometric reverberation
mapping observations \citep{Ka00}, and the sampling rates of the
lines are around five times per year. Light curves of broad UV
lines C~{\sc iv} $\lambda 1549$ and Ly$\alpha$ $\lambda 1216$ are
from International Ultraviolet Explorer observations \citep{Pa03}.
The sampling rates of the Ly$\alpha$ and C~{\sc iv} lines are
around six times per year. The data numbers and durations of light
curves are listed in Table 3. All the light curves are presented
in Fig. 5.
\begin{figure}
\begin{center}
\includegraphics[angle=-90,scale=0.30]{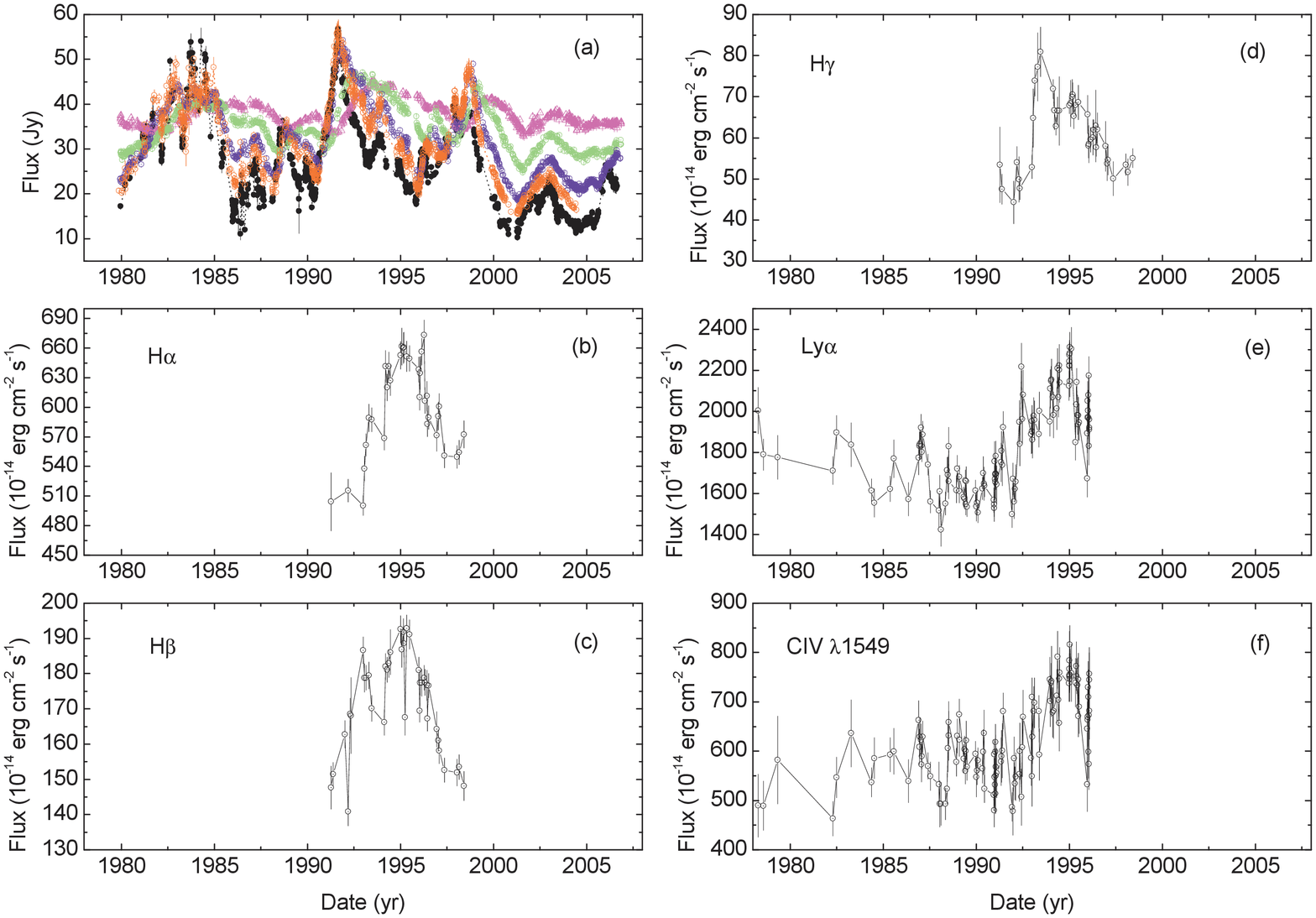}
\end{center}
 \caption{Light curves of 3C 273. (a) radio light curves: black color denotes 37 GHz light curve, red color 22 GHz one, blue color 15 GHz one,
green color 8 GHz one, and magenta color 5 GHz one. (b) H$\alpha$
light curve. (c) H$\beta$ light curve. (d) H$\gamma$ light curve.
(e) Ly$\alpha$ light curve. (f) C~{\sc iv} $\lambda 1549$ light
curve.}
  \label{fig5}
\end{figure}

\begin{table}
\centering
\begin{minipage}{80mm}

\caption{Fractional variability of 3C 273 \label{tbl-3}}

\begin{tabular}{ccccc}

\hline\hline

Component & $F_{\rm{var}}$ & $\sigma_{\rm{F_{\rm{var}}}}$ & $N$ & $D$ (yr)\\
(1)&(2)&(3)&(4)&(5)\\

\hline

H$\alpha$                &0.08   &0.01  & 34   & 7.13   \\
H$\beta$                 &0.08   &0.01  & 39  &  7.13  \\
H$\gamma$                &0.12  &0.02   & 39   & 7.13   \\
C~{\sc iv} $\lambda 1549$      &0.12  & 0.01 & 119  & 17.79    \\
Ly$\alpha$ $\lambda 1216$&0.11   &0.01 & 119  & 17.79  \\
5 GHz                    &0.072  &0.002&784    & 26.90 \\
8 GHz                    &0.143 &0.003 &1076   &26.89 \\
15 GHz                   &0.259 &0.006 &1063   &26.88  \\
22 GHz                   &0.286 &0.006 &1092   & 24.59 \\
37 GHz                   &0.372 &0.008 &1247   & 26.67 \\

\hline
\end{tabular}
\\Notes: Columns are same as Table 1.
\end{minipage}
\end{table}

The radio light curves after $\sim 1980$ are cross-correlated with
the broad-line light curves. We performed 10,000 runs of Monte
Carlo simulations, and the CCCDs are listed in Fig. 6. These time
lags and uncertainties are listed in Table 4. All the CCCDs have a
negative time lag (see Fig. 6), which indicates that the radio
variations lead the broad-line variations. These CCCDs vary for
different radio and broad-line light curves. The CCCDs between the
H$\gamma$ line and radio variations seem to have similar profiles,
and in general each of them is within a narrower interval. These
similar and narrower CCCDs indicate good cross-correlations
between the H$\gamma$ broad-line and radio jet emission
variations.
\begin{figure*}
 \includegraphics[angle=-90,scale=0.20]{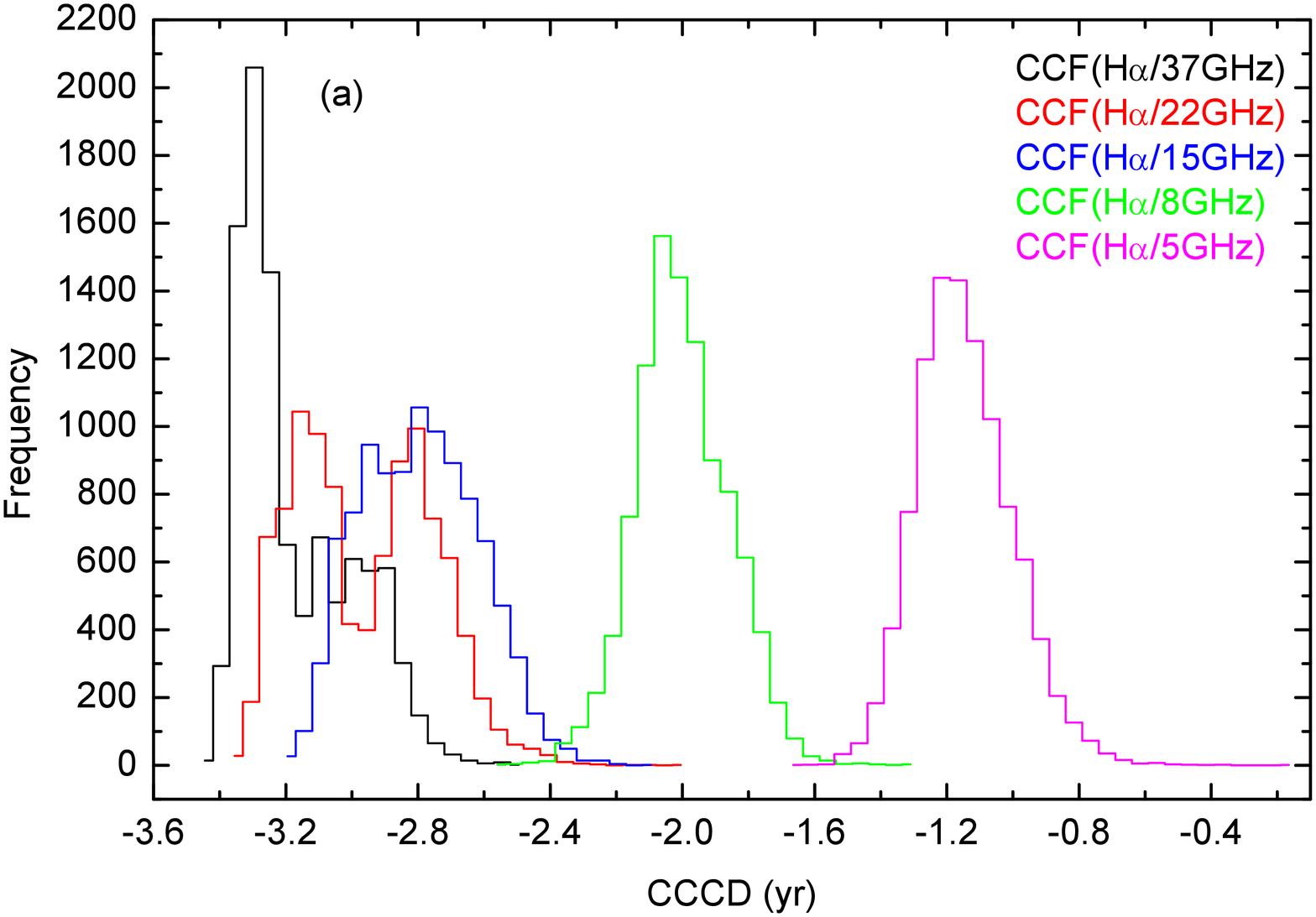}
 \includegraphics[angle=-90,scale=0.20]{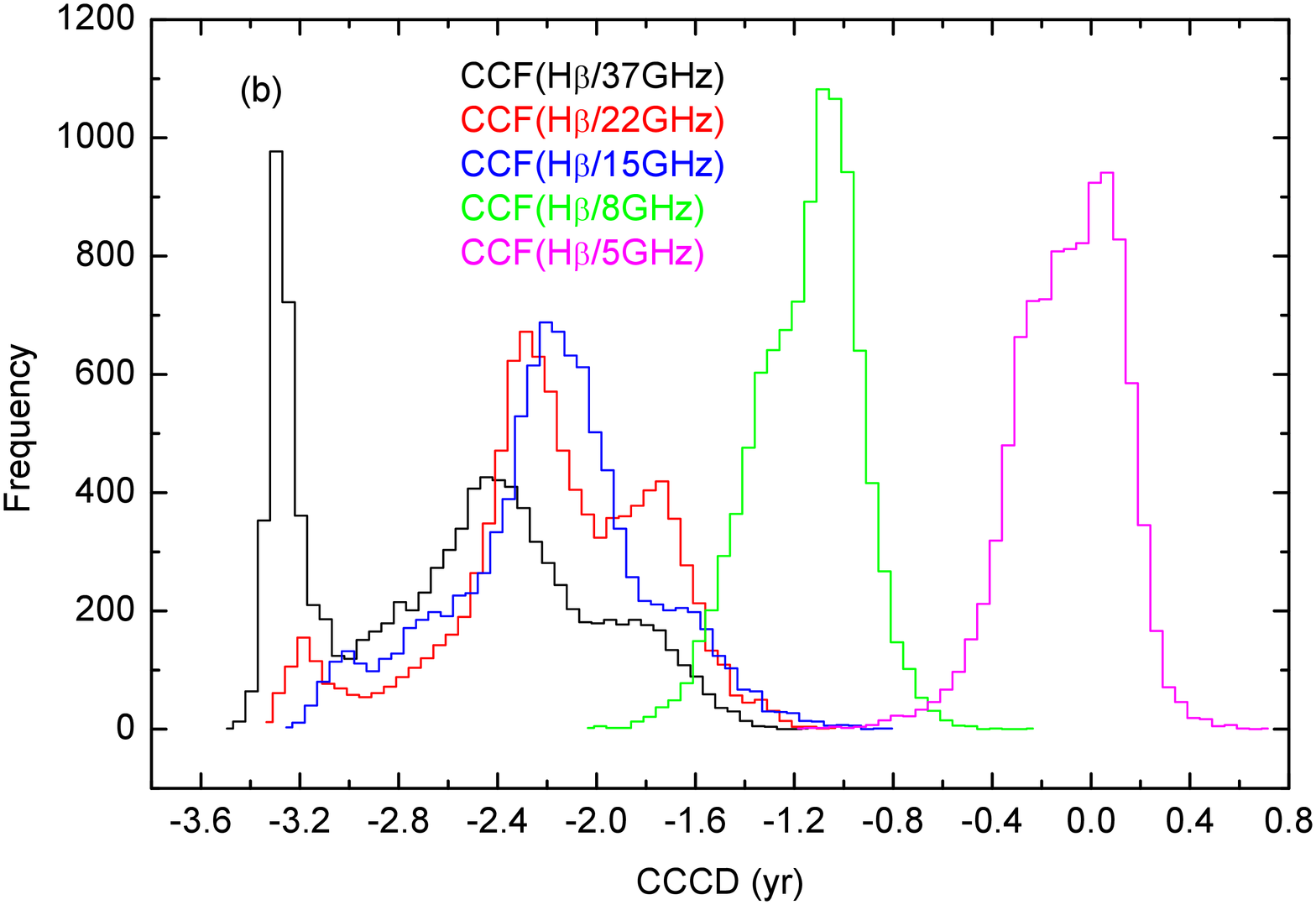}
  \includegraphics[angle=-90,scale=0.20]{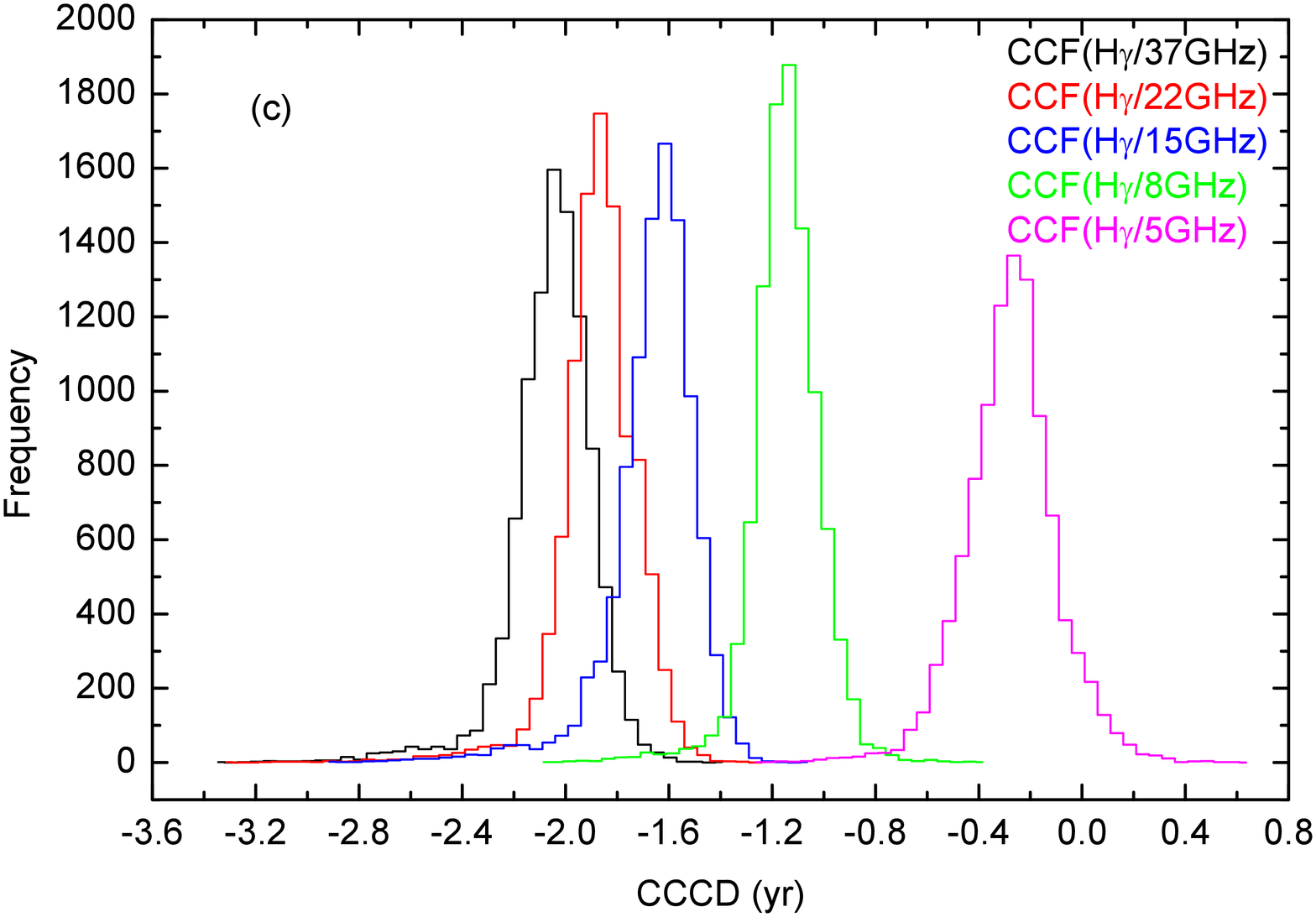}
 \includegraphics[angle=-90,scale=0.20]{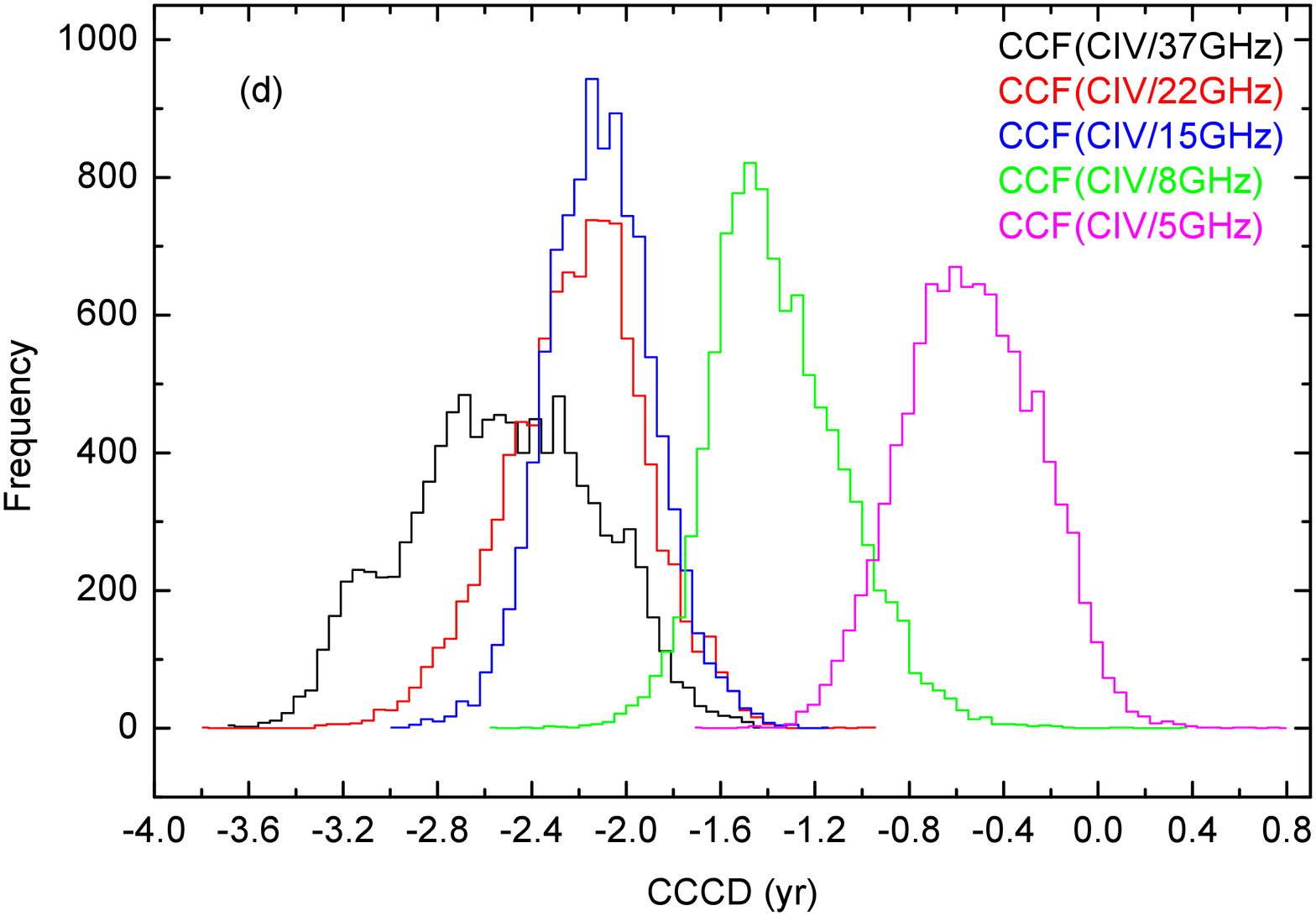}
  \includegraphics[angle=-90,scale=0.20]{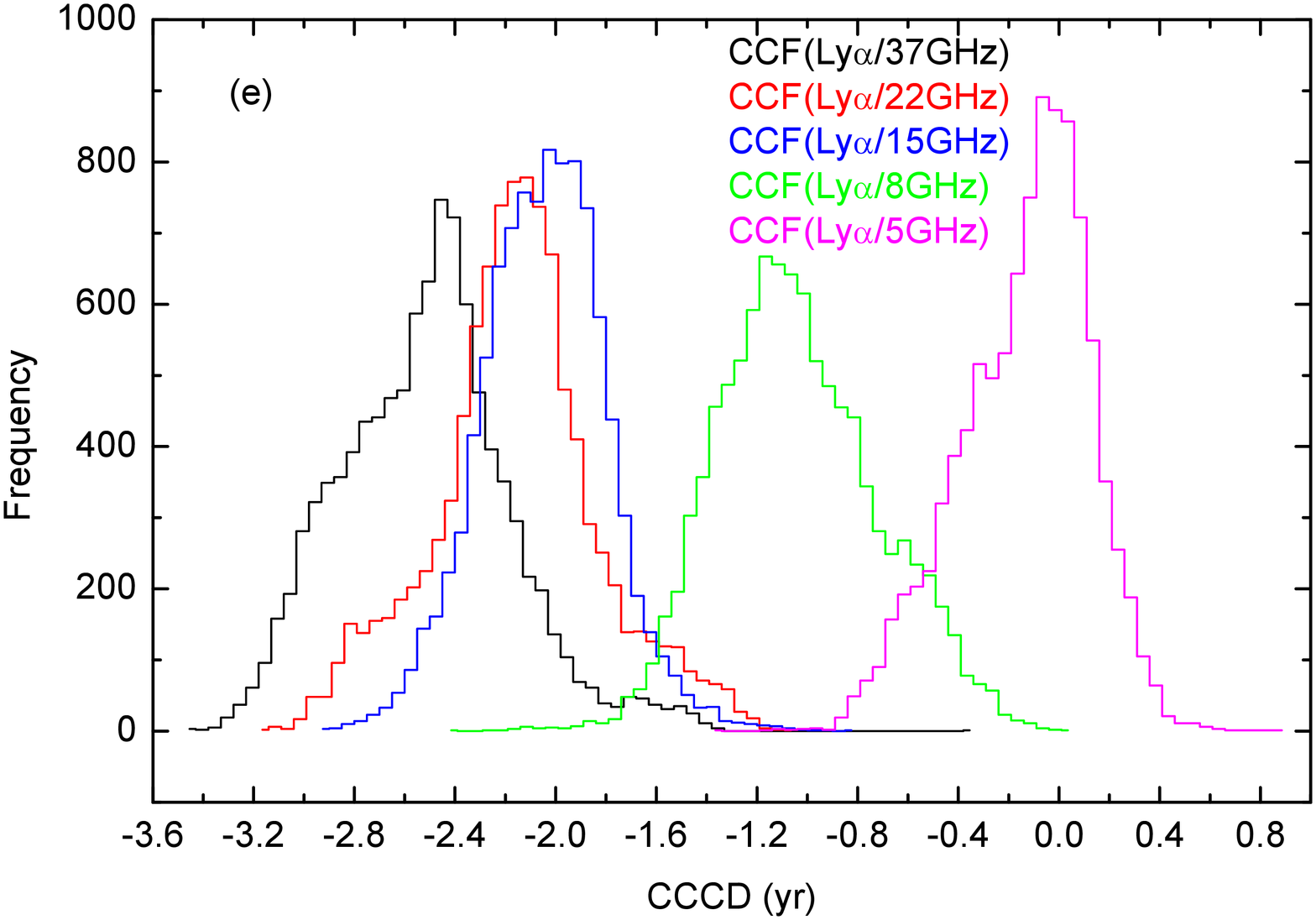}
 \caption{CCCDs obtained with the FR/RSS method in Monte Carlo simulations of 10,000 runs for 3C 273.
 (a) H$\alpha$ line, (b) H$\beta$ line, (c) H$\gamma$ line, (d) C~{\sc iv} $\lambda 1549$ line, and
 (e) Ly$\alpha$ line relative to radio emission variations.}
  \label{fig6}
\end{figure*}

\begin{table*}
\centering
\begin{minipage}{100mm}
\caption{Time lags between broad-lines and radio emission for 3C
273 \label{tbl-4}}

\begin{tabular}{cccccc}

\hline\hline

Lines & 5 GHz & 8 GHz & 15 GHz & 22 GHz & 37 GHz \\
(1)&(2)&(3)&(4)&(5)&(6)\\

\hline

H$\alpha$&$-1.15\pm 0.15$&$-2.00\pm 0.14$&$-2.80\pm 0.18$&$-2.96\pm0.21$&$-3.17\pm 0.17$\\
H$\beta$&$-0.09\pm 0.21$&$-1.15\pm 0.21$&$-2.18\pm 0.40$&$-2.18\pm 0.42$&$-2.60\pm 0.53$\\
H$\gamma$&$-0.28\pm 0.18$&$-1.15\pm 0.14$&$-1.66\pm 0.17$&$-1.87\pm 0.16$ &$-2.05\pm 0.17$\\

C~{\sc iv}&$-0.54\pm 0.28$&$-1.35\pm 0.28$&$-2.11\pm 0.23$&$-2.21\pm 0.30$&$-2.53\pm 0.39$\\
Ly$\alpha$&$-0.14\pm 0.26$&$-1.04\pm 0.32$&$-2.05\pm 0.25$&$-2.16\pm 0.33$&$-2.52\pm 0.34$\\

\hline
\end{tabular}
\\Notes: Signs of time lags are defined as
$\tau_{\rm{cent}}= t_{\rm{radio}}-t_{\rm{line}}$, in units of yr.
\end{minipage}
\end{table*}

The broadband spectral energy distributions of 3C 273 constrain a
Doppler factor $\delta =6.5$ \citep{Gh98}. These radio variations
at high frequencies lead those of low frequencies (see Paper I).
The time lags between the 5, 8, 15, 22, and 37 GHz variations can
be explained by the radiation cooling effect of relativistic
electrons with $\delta =6.5$ (see Fig. 6 in Paper I). The Doppler
factor $\delta \ge 1.90$ is given by \citet{Xi04} using the
minimum timescale of variations at the optical band. The radiative
cooling can match the time lags between these radio variations for
$\delta =$3.5--6.5 (see Fig. 7). Thus we take $\delta =$3.5--6.5
with $\delta=\sqrt{1-\beta^2}/(1-\beta \cos \alpha)$ as a
constraint in equations (8), (13), and (14). As in Paper I, we
take $\alpha=12^{\circ}$--$21^{\circ}$ and $\beta=$0.9--0.995.

The sizes of BLRs for the Balmer lines are somehow controversial
for 3C 273. \citet{Pa05} think that the H$\alpha$, H$\beta$ and
H$\gamma$ lags relative to the UV continuum are more reliable than
those relative to the optical continuum. The UV continuum is more
appropriate than the optical continuum as the ionizing continuum
of the Balmer lines. The time lag of the H$\gamma$ line variations
relative to the 37 GHz variations is $\tau_{\rm{ob}}=-2.05\pm
0.17$ yr (see Table 4). The H$\gamma$ line has a BLR size of
$R_{\rm{BLR}}=2.85\pm 0.32$ lt-yr relative to the UV continuum
\citep{Pa05}. Thus we get $R_{\rm{jet}}=8.63\pm 1.16$ lt-yr from
Monte Carlo simulations based on equation (14) (see Table 5). This
location of $R_{\rm{jet}}=2.65\pm 0.36$ pc is outside the BLR.
Another possible choices of $R_{\rm{BLR}}$ and $\tau_{\rm{ob}}$
are the averages of the two quantities for the three Balmer lines.
The average time lag is $\tau_{\rm{ob}}=-2.61\pm 0.17$ yr between
the variations of the Balmer lines and the 37 GHz emission. The
average BLR size is $R_{\rm{BLR}}=2.70\pm 0.13$ lt-yr for the
Balmer lines relative to the UV continuum \citep{Pa05}. We have
$R_{\rm{jet}}=3.27\pm 0.67$ lt-yr, i.e., $R_{\rm{jet}}=1.00\pm
0.21$ pc from the central engine. This position is around the BLR.
These two estimated $R_{\rm{jet}}$ are at the pc-scale distance
from the central engine, and the emitting regions on the pc-scales
in the jet are difficult to be resolved in imaging observations.

\begin{figure}
 \begin{center}
 \includegraphics[angle=-90,scale=0.25]{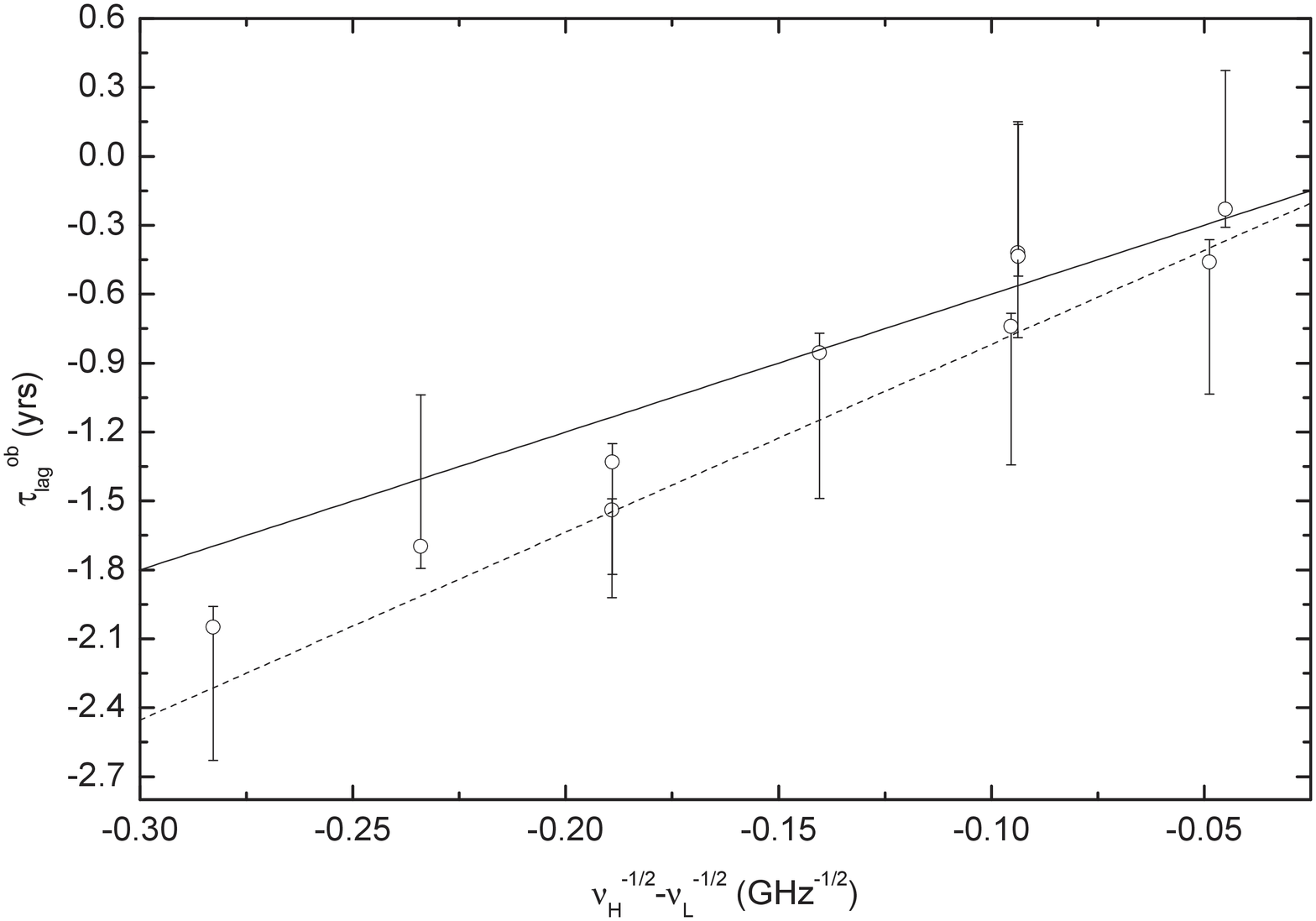}
 \end{center}
 \caption{Same as Fig. 6 in Paper I. Time lags $\tau^{\rm{ob}}_{\rm{lag}}=t_{\rm{\nu_{H}}}-t_{\rm{\nu_{L}}}$ versus frequency
 differences $\nu^{-1/2}_{\rm{H}}-\nu^{-1/2}_{\rm{L}}$. Solid line is the expectation from the radiative cooling with $\delta =6.5$.
 Dashed line is for $\delta =3.5$.}
  \label{fig7}
\end{figure}

\section{Variability Amplitude}
The root-mean-square fractional variability amplitude is used to
measure the variability of a light curve, and the fractional
variability amplitude $F_{\rm{var}}$ is \citep{Ro97}
\begin{equation}
F_{\rm{var}}=\sqrt{\frac{S^2-\langle\sigma^2_{\rm{err}}\rangle}{\langle
F\rangle^2}},
\end{equation}
where $\langle F \rangle$ is the mean flux, $S^2$ the variance,
and $\langle \sigma^2_{\rm{err}} \rangle$ the measured mean square
error. The error of $F_{\rm{var}}$ is \citep{Ed02}
\begin{equation}
\sigma_{F_{\rm{var}}}=\frac{1}{F_{\rm{var}}}\sqrt{\frac{1}{2N}}\frac{S^2}{\langle
F\rangle^2},
\end{equation}
where $N$ is the number of data points in the light curve.
$F_{\rm{var}}$ and $\sigma_{F_{\rm{var}}}$ are estimated for all
the light curves for 3C 120 and 3C 273 (see Tables 1 and 3). For
3C 273, the broad lines and the 5 and 8 GHz emission have a
comparable $F_{\rm{var}}$, i.e., a comparable variability. The 15,
22 and 37 GHz emission also have a comparable $F_{\rm{var}}$.

For 3C 120, the H$\beta$ line light curves have the same
$F_{\rm{var}}$ at three periods. There are also comparable
$F_{\rm{var}}$ for the H$\beta$ line total light curve, the
H$\gamma$ line, the He II $\lambda 4686$ line, and the 15 GHz
emission. Three components in the modified 15 GHz light curve are
compared to the moved light curves of the H$\gamma$ and He II
$\lambda 4686$ lines in \citet{Ko14}, the H$\beta$ line in
\citet{Gr12}, and the H$\beta$ line in \citet{Nu12} (see Fig. 8).
The H$\gamma$ and He II $\lambda 4686$ line light curves have
similar profiles to Outburst I. There is a good correspondence
between Outburst II and the H$\beta$ line light curve in
\citet{Gr12}. Event A shows a correspondence to the H$\beta$ line
light curve in \citet{Nu12}. Comparisons of the line light curves
to Outbursts I and II and Event A show correspondences between the
line and radio variations. The moved times in Fig. 8 are
consistent with the time lags listed in Table 2, except for the
H$\beta$ light curve and Event A, moved by 0.55 yr. This H$\beta$
light curve and Event A are at low states (see Fig. 3). Event A is
likely produced by a weaker radio knot with a lower velocity, and
the knot needs more times to travel from the central engine to the
radio emitting region. Equation (8) shows that the time lag
$\tau_{\rm{ob}}$ increases as the velocity $\beta_{\rm{a}}$
decreases if the radio emitting region is roughly around a
position $R_{\rm{jet}}$. It is natural that the moved times of
0.55 yr for this H$\beta$ light curve and Event A is larger than
the time lag of $\tau_{\rm{cent}}=0.35 \pm 0.04$ yr for the total
H$\beta$ light curve. Thus it should be reliable for 3C 120 that
the line light curves correspond to Outbursts I and II and Event A
presented in Fig. 8. The moved light curves of He II $\lambda
4686$ and H$\gamma$ lines have the same $F_{\rm{var}}$ consistent
with that of Outburst I (see Table 1). Outburst II has a
$F_{\rm{var}}$ consistent with that of the moved H$\beta$ light
curve in \citet{Gr12} (see Table 1). Event A has a $F_{\rm{var}}$
smaller than that of the moved H$\beta$ light curve in
\citet{Nu12} (see Table 1). These indicate that the variability
amplitudes do not have inevitable relations with the
correspondences (cross-correlations) between the broad-line and
radio jet emission variations.
\begin{figure}
 \begin{center}
 \includegraphics[angle=-90,scale=0.3]{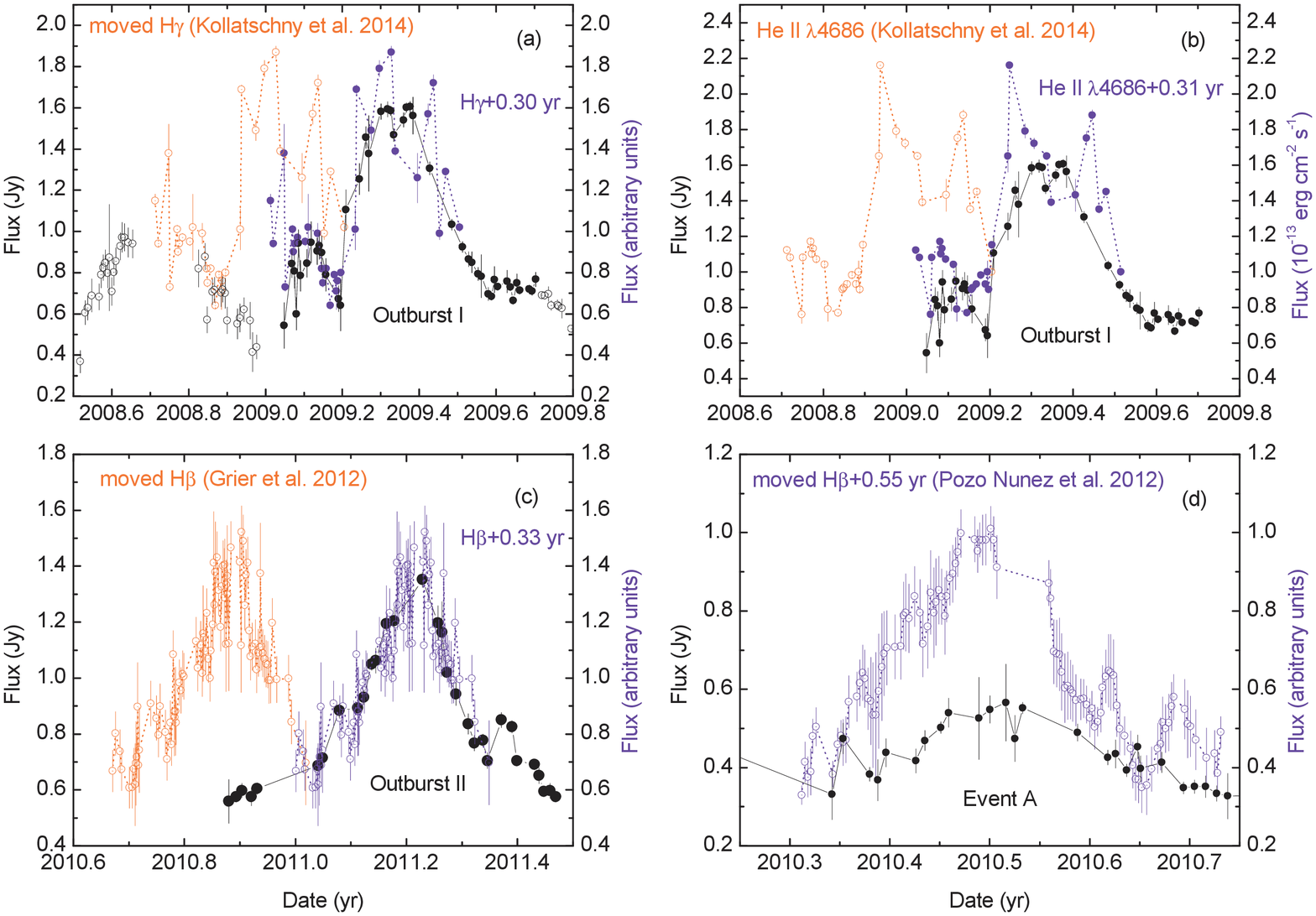}
 \end{center}
 \caption{3C 120: comparisons of 15 GHz components to the line light curves moved vertically and/or horizontally.
 The black solid circles are three components in the 15 GHz light curve, and the blue ones
 the moved line light curves.}
  \label{fig8}

\end{figure}

\section{DISCUSSION AND CONCLUSIONS}

Another configuration of BLR is a disk-like structure \citep[see
e.g. ][]{Ko14,Nu14}. The disk-like BLR has a ratio of height to
radius $C=H/r$. For a thin ring with a height $H$ in the range of
$r\rightarrow r+dr$, it has a covering factor of
$df_{\rm{cov}}(r)=n_{\rm{c}}(r)\sigma_{\rm{c}}(r) dr$ and a volume
$dv=2\pi r H dr$. The emissivity of the thin ring within
$r\rightarrow r+dr$ due to the photoionization of ultraviolet
luminosity $L_{\rm{UV}}$ is
\begin{equation}
j_{\rm{BLR}}(r)=\frac{L_{\rm{UV}}df_{\rm{cov}}(r)}{4\pi dv }
=\frac{L_{\rm{UV}} n_{\rm{c}}(r) \sigma_{\rm{c}}(r)} {8\pi^2 C
r^2}\propto \frac{n_{\rm{c}}(r) \sigma_{\rm{c}}(r)}{r^2}.
\end{equation}
Weighted averaging equation (8) over the whole BLR with the
differential BLR flux within the range of $r\rightarrow r+dr$, and
we have
\begin{equation}
R_{\rm{jet}}=\frac{\beta_{\rm{a}}}{\sin \alpha} \left( \frac{\int
^{R_{\rm{BLR,out}}}_{R_{\rm{BLR,in}}}j_{\rm{BLR}}(r)rdr}{\int
^{R_{\rm{BLR,out}}}_{R_{\rm{BLR,in}}}j_{\rm{BLR}}(r)dr}+\frac{\langle\tau_{\rm{ob}}
\rangle}{1+z} c \right).
\end{equation}
If $n_{\rm{c}}(r)$ and $\sigma_{\rm{c}}(r)$ have pow-law profiles,
equation (18) will have the same expression as equation (13). For
the disk-like BLR, the pow-law indexes $q$ and $p$ do not have the
values similar to those derived from fitting the observed line
light curves on the basis of photoionization calculations of a
large number of clouds for the spherical BLR in \citet{Ka99}.
Recently, \citet{Kh15} investigated orbital motion of spherical,
pressure-confined clouds in the BLR of AGNs, and found that a
disk-like configuration is more plausible for the distribution of
the BLR clouds. For a pressure-confined cloud,
$r_{\rm{c}}(r)\varpropto P_{\rm{gas}}^{-1/3}$, where
$P_{\rm{gas}}$ is the intercloud gas pressure and
$P_{\rm{gas}}\varpropto r^{-5/2}$ \citep{Kh15}. Therefore, the
cloud has a radius profile of $r_{\rm{c}}(r)\varpropto r^{5/6}$.
This profile is similar to that one $r_{\rm{c}}(r)\varpropto
r^{1/3}$ in \citet{Ka99}. A radial surface line emissivity profile
of $F(r)\propto r^{-1}$ is assumed (in units of $\rm{erg \/\
s^{-1} \/\ cm^{-2}}$), and it is a fair approximation to the
expected radial distribution derived from photoionization
calculations for several of the commonly observed UV and optical
emission lines \citep{Go12}. Thus, there is a differential
broad-line flux $d F_{\rm{BLR}}(r) \varpropto r^{-2}dr$ in the
range of $r\rightarrow r+dr$ for this profile. In the case of $d
F_{\rm{BLR}}(r) \varpropto r^{-2}dr$, we have
\begin{equation}
\begin{split}
R_{\rm{jet}}&=\frac{\beta_{\rm{a}}}{\sin \alpha} \left( \frac{\int
^{R_{\rm{BLR,out}}}_{R_{\rm{BLR,in}}}r d F_{\rm{BLR}}(r)}{\int
^{R_{\rm{BLR,out}}}_{R_{\rm{BLR,in}}}d
F_{\rm{BLR}}(r)}+\frac{\langle\tau_{\rm{ob}} \rangle}{1+z}
c\right)   \\
&=\frac{\beta_{\rm{a}}}{\sin \alpha} \left( \frac{\int
^{R_{\rm{BLR,out}}}_{R_{\rm{BLR,in}}} r^{-1} dr}{\int
^{R_{\rm{BLR,out}}}_{R_{\rm{BLR,in}}} r^{-2}
dr}+\frac{\langle\tau_{\rm{ob}} \rangle}{1+z} c \right)\\
&=\frac{\beta_{\rm{a}}}{\sin \alpha} \left( \frac{\ln
R_{\rm{BLR,out}}-\ln
R_{\rm{BLR,in}}}{R_{\rm{BLR,in}}^{-1}-R_{\rm{BLR,out}}^{-1}}+\frac{\langle\tau_{\rm{ob}}
\rangle}{1+z} c \right).
\end{split}
\end{equation}
At the same time, $d F_{\rm{BLR}}(r) \varpropto j_{\rm{BLR}}(r) dr
\varpropto n_{\rm{c}}(r) \sigma_{\rm{c}}(r) r^{-2}dr \varpropto
r_{\rm{c}}^2 r^{-p} r^{-2}dr=r^{5/3-p-2}dr$
($r_{\rm{c}}(r)\varpropto r^{5/6}$). So, $p=5/3$ for the disk-like
BLR. In the spherical BLR, $p=3/2$. The cloud number density
profile of the spherical BLR is consistent with that of the
disk-like BLR ($n_{\rm{c}}(r)\varpropto r^{-p}$ with comparable
$p$ for the two kinds of BLRs).

The disk-like BLR geometry of 3C 120 has been established in
\citet{Ko14} and \citet{Nu14}. Then $R_{\rm{jet}}$ will be
re-calculated on the basis of equation (19). Based on
$\tau_{\rm{cent}}$, $R_{\rm{BLR,in}}$ and $R_{\rm{BLR,out}}$ of
the H$\beta$, H$\gamma$ and He II $\lambda 4686$ lines,
$\beta_{\rm{a}}=4.0\pm 0.2$, $\alpha = 20.5\pm 1.8^{\circ}$, and
equation (19), averages of $R_{\rm{jet}}$ are derived from Monte
Carlo simulations. These values of $R_{\rm{jet}}$ estimated under
the disk-like BLR are consistent with those under the spherical
BLR for 3C 120 (see Table 2). Equation (8) is based on a simple
spherical shell or ring with a zero-thickness, and we also
re-calculate $R_{\rm{jet}}$ with equation (8). The estimated
results are presented in Table 2. These values are consistent with
those estimated from equations (14) and (19). Therefore, the four
BLRs with different configurations result in some negligible
influences on $R_{\rm{jet}}$ for 3C 120. Modelling photometric
reverberation data favors a nearly face-on disk-like BLR geometry
with an inclination $i = 10\pm 4^{\circ}$ and an extension from 22
to 28 light-days \citep{Nu14}. If the viewing angle $\alpha=10\pm
4^{\circ}$ is adopted (and $R_{\rm{BLR}}$ spans from 22 to 28
light-days), $R_{\rm{jet}}$ is larger by a factor 2.0 than that
value for $\alpha = 20.5\pm 1.8^{\circ}$ and
$R_{\rm{BLR}}=0.075\pm 0.003$ lt-yr. Thus the influence of the
inclination of the disk-like BLR is more significant than that of
the BLR configurations on $R_{\rm{jet}}$. This is also indicated
by the dependence of $R_{\rm{jet}}$ on $\alpha$ scaling as
$R_{\rm{jet}}\propto 1/\sin \alpha$. For the spherical BLR, the
cloud density profile also has a $n_{\rm{c}}\propto r^{-p}$ with a
possible value of $p=2$ \citep{Ka99}. $q=1/2$ or $ q=1/3$ are
possible values of the best models in \citet{Ka99}. Thus we
consider the four combinations of $q=1/3,1/2$ and $p=3/2,2$ in
equation (13) for 3C 273. The estimated values of $R_{\rm{jet}}$
are presented in Table 5. It is obvious that the different
combinations of $q$ and $p$ only have weaker influences on
$R_{\rm{jet}}$, and the influences are negligible for 3C 273.
\begin{table*}
\centering
\begin{minipage}{130mm}
\caption{Influences of $q$ and
$p$ on $R_{\rm{jet}}$ for 3C 273\label{tbl-5}}

\begin{tabular}{ccccccc}
\hline
\hline

Lines & $R_{\rm{BLR}}$ (lt-yr) & $\tau_{\rm{cent}}$(yr) & \multicolumn{4}{c}{$R_{\rm{jet}}$ (pc)}   \\
\cline{4-7}

 & & &$\left(1/3,3/2\right)$&$\left(1/3,2\right)$&$\left(1/2,3/2\right)$&$\left(1/2,2\right)$ \\
(1)&(2)&(3)&(4)&(5)&(6)&(7) \\
\hline

Balmer&$2.70^{+0.13}_{-0.13}$&$-2.61\pm 0.17$&$1.00\pm 0.21$ &$1.00\pm 0.20$&$1.00\pm 0.21$&$1.00\pm 0.20$\\
H$\gamma$&$2.85^{+0.32}_{-0.32}$&$-2.05\pm 0.17$&$2.65\pm 0.36$&$2.63\pm 0.35$&$2.66\pm 0.36$ &$2.64\pm 0.35$\\
\hline

\end{tabular}
\\Notes: Column 1: line names; Column 2: BLR sizes:
$R_{\rm{BLR,in}}=R_{\rm{BLR}}-\sigma_{\rm{R_{\rm{BLR}}}}$ and
$R_{\rm{BLR,out}}=R_{\rm{BLR}}+\sigma_{\rm{R_{\rm{BLR}}}}$; Column
3: Time lags defined as $\tau_{\rm{cent}}=
t_{\rm{radio}}-t_{\rm{line}}$ between broad-lines and radio
emission; Columns 4--7: $R_{\rm{jet}}$ estimated with equation
(13) for different combinations of $(q,p)$. Balmer lines mean that
$R_{\rm{BLR}}$ and $\tau_{\rm{cent}}$ are the averages of
H$\alpha$, H$\beta$, and H$\gamma$ lines.

\end{minipage}
\end{table*}

The emitting regions of radio outbursts are at the pc-scales from
the central engines for 3C 120 and 3C 273 (see Tables 2 and 5).
These regions may have an important impact on the gamma rays in 3C
120 and 3C 273, because that the gamma-ray emitting position
relative to the BLR plays an important role in the gamma-ray
emission from the jet
\citep[e.g.][]{Si94,Wa00,LB06,Li08,Si08,Ta08,BLM09,Ta09,Le14a}.
Recently, \citet{Ma14} investigated the cross-correlations between
these light curves of the brightest detected blazars from the
first 3 years of the mission of \textit{Fermi}-LAT and 4 years of
15 GHz observations from the OVRO 40 m monitoring program. They
found for four sources that the radio variations lag the gamma-ray
variations, suggesting that the gamma-ray emission originate
upstream of the radio emission, i.e., $R_{\rm{\gamma}}\la
R_{\rm{jet}}$. The constraint of $R_{\rm{\gamma}}\la R_{\rm{jet}}$
was also suggested in some researches
\citep[e.g.][]{De94,Jo01,Ko09,Si09,AAA10a}. Thus, we have
$R_{\rm{\gamma}}\la R_{\rm{jet}}\simeq$ 1.0--1.5 pc for 3C 120 and
$R_{\rm{\gamma }}\la$1.0--2.6 pc for 3C 273. If we have known the
relative sizes of $R_{\rm{jet}}$ to $R_{\rm{\gamma}}$ and
$R_{\rm{BLR}}$, the relative size of $R_{\rm{\gamma}}$ to
$R_{\rm{BLR}}$ would be constrained. The relative size of
$R_{\rm{\gamma}}$ to $R_{\rm{BLR}}$ may be determined by a time
lag between gamma-ray and broad-line variations if there is
correlation. For 3C 120, $R_{\rm{jet}}\gg R_{\rm{BLR}}$, and then
it is possible $R_{\rm{\gamma}}\gg R_{\rm{BLR}}$, which limits the
EC component of gamma rays to be negligible compared to the SSC
one. \citet{Ta15} derived a ratio of EC to SSC luminosity of
$\sim$ 0.1 for 3C 120. The dominant SSC component deduced here is
consistent with \citet{Ta15}. For 3C 273, $R_{\rm{jet}}$ is
slightly larger than $R_{\rm{BLR}}$. $R_{\rm{\gamma}}$ may be
around or smaller than $R_{\rm{BLR}}$, and then the SSC component
may be comparable to the EC one of gamma rays, i.e., the EC
component is not negligible. In fact, the gamma-ray emitting
position is complex relative to the BLR. For example, most of the
time the gamma-ray emitting region is inside the BLR, but during
some epoches the emitting region could drift outside the BLR.
\citet{Fo11} proposed the very first idea, and recently
\citet{Gh13} found a very clear case with multi-wavelength
coverage. These will increase the complexity of the gamma-ray
emission.

In this paper, we first derived a new formula under the spherical
shell BLR with a zero thickness, and the formula connects
$R_{\rm{jet}}$, $R_{\rm{BLR}}$, $\tau_{\rm{ob}}$,
$\beta_{\rm{a}}$, and $\alpha$. The new formula is the same as
that obtained under the ring BLR with a zero thickness (see Paper
I). Second, we derived new formulae under the spherical BLR, a
classical configuration, with cloud number density and radius
radial profiles. The new formulae for the spherical BLR are
applied to broad-line radio-loud \textit{Fermi}-LAT AGNs 3C 120
and 3C 273. We analyzed the cross-correlations between broad-line
and radio jet emission variations on the basis of the
model-independent FR/RSS method and/or the ZDCF method. For 3C
120, a newly published paper presents H$\beta$, H$\gamma$ and He
II $\lambda 4686$ line new data in reverberation mapping
observations. Combined with the data sets of H$\beta$ line in
other two papers, a longer light curve is used to cross-correlate
with the 15 GHz light curve. The 15 GHz radio variations lag the
broad-line H$\beta$, H$\gamma$ and He II $\lambda 4686$
variations, i.e., $\tau_{\rm{ob}}>0$, and
$R_{\rm{jet}}\simeq$1.1--1.5 pc are estimated on the basis of the
spherical BLR, the disk-like BLR, and the spherical shell and/or
ring BLR (see Table 2). $R_{\rm{jet}}\gg R_{\rm{BLR}}$ for 3C 120,
and this position far away from the central engine may have
important influences on the gamma-ray emission. The radio
variations lead the broad-line variations, i.e.,
$\tau_{\rm{ob}}<0$ for 3C 273. $R_{\rm{jet}}\simeq$1.0--2.6 pc are
derived from the negative time lags for the spherical BLR. For 3C
273, $R_{\rm{jet}}\ga R_{\rm{BLR}}$, and we have
$R_{\rm{\gamma}}\la R_{\rm{jet}}\simeq$ 1.0--2.6 pc. The gamma-ray
flares detected with $\textit{Fermi}$-LAT set a limit of
$R_{\rm{\gamma}}<$ 1.6 pc for 3C 273 \citep{Ra13}. The limit is
marginally consistent with our constraint of $R_{\rm{\gamma
}}\la$1.0--2.6 pc. This agreement indicates the reliability of the
method used to estimate $R_{\rm{jet}}$. For 3C 120,
$R_{\rm{\gamma}}\la R_{\rm{jet}}\simeq$ 1.0--1.5 pc. The cloud
number density and radius radial profiles of the BLR have
negligible influences on $R_{\rm{jet}}$ (see Table 5), and also
the BLR configurations do (see Table 2). The inclination of the
disk-like BLR will have a significant influence on $R_{\rm{jet}}$
(the viewing angle $\alpha$ is same as this inclination for the
assumption that the jet axis is  perpendicular to the plane of
this BLR).

The black hole mass is of the order of $10^{7} M_{\rm{\odot}}$ in
3C 120 \citep{Pe98a,Pe04,Gr12,Nu12}. Recently, \citet{Ko14} and
\citet{Nu14} derived larger masses of the order of $10^{8}
M_{\rm{\odot}}$ for 3C 120. 3C 120 has $R_{\rm{jet}}=$1.1--1.5 pc
and a mass of the order of $10^{8} M_{\rm{\odot}}$. 3C 273 has
$R_{\rm{jet}}=$1.0--2.6 pc and a mass of the order of $10^{9}
M_{\rm{\odot}}$ \citep{Pa05}. The radio emitting positions do not
seem to scale with the masses of the central black holes for the
two broad-line radio-loud \textit{Fermi}-LAT AGNs. Until the
moment there is no evidence that the positions of the emitting
regions in the jets of AGNs scale with the masses of the central
black holes, though this scaling relation (if present) will be
important to the jet production and energy dissipation mechanisms
in AGNs. Also, this scaling relation will be important with regard
to jet-enhanced disk accretion in AGNs \citep{Jo07a,Jo07b}. In the
future, the radio and gamma-ray emitting positions will be needed
for more AGNs.

\section*{Acknowledgments}
We are grateful to the anonymous referee for important suggestions
and acting assistant editor Mr Kulvinder Singh Chadha for helpful
comments leading to significant improvement of this paper. HTL
thanks the National Natural Science Foundation of China (NSFC;
Grant 11273052) for financial support. JMB acknowledges the
support of the NSFC (Grant 11133006). HTL also thanks the
financial support of the Youth Innovation Promotion Association,
CAS and the project of the Training Programme for the Talents of
West Light Foundation, CAS. This research has made use of data
from the OVRO 40 m monitoring program which is supported in part
by NASA grants NNX08AW31G and NNX11A043G, and NSF grants
AST-0808050 and AST-1109911.

\label{lastpage}

\end{document}